\documentclass{cernrep} 
\usepackage{texnames}
\usepackage[T1]{fontenc}
\usepackage[bookmarks, colorlinks=true, linktoc=page, pdftex, linkcolor=blue, citecolor=blue, urlcolor=blue]{hyperref}
\usepackage{graphicx,latexsym}
\usepackage{subcaption} 

\usepackage{tcolorbox}

\usepackage{wrapfig}
\usepackage{mdframed}
\usepackage{enumitem}
\usepackage{afterpage}
\usepackage{lipsum}
\usepackage{array}
\usepackage{xcolor, soul}
\usepackage{colortbl}

\newcommand{\bit}{\begin{Itemize}}
\newcommand{\eit}{\end{Itemize}}

\usepackage{fancyhdr}
\fancyhfoffset{4 mm}
\fancypagestyle{ARTTITLE}{%
\fancyhf{} 
\lhead{\small{Proceedings of the CERN--Accelerator--School course: \it{Introduction to Accelerator Physics}, courses 2019 and beyond}}
\lfoot{Available online at \url{https://cas.web.cern.ch/previous-schools}}
\rfoot{\thepage\hspace*{3mm}}

}
\begin{document}
\title{Introduction to Special Relativitiy}

\author{E.~Gianfelice-Wendt}
\institute{Fermilab, Batavia IL, US}


\begin{abstract}

The goal of this lecture is to introduce the student to the theory of Special Relativity.

\noindent
Not to overload the content with mathematics, the author will stick to the simplest cases; in particular only reference frames  using
Cartesian coordinates and 
translating along the common $x$-axis as in Fig.~\ref{eliana-rel-f1} will be used.

The general expressions will be quoted
or may be found in the cited literature.

\end{abstract}

\maketitle 
\thispagestyle{ARTTITLE}

\section{Introduction}
In the second half of the
XIX century Maxwell had summarized all known electromagnetic phenomena
in four partial differential equations for  electric and magnetic fields. These equations contain a numerical constant, $c$,
which has the dimension of a velocity and the value of the speed of light in vacuum. 
Far from the sources, the Maxwell equations contains also the wave equation
$$
\left[ \nabla^2 - \frac{1}{c^2}\frac{\partial^2}{\partial t^2}\right]\Phi  = 0 
$$
where the constant $c$ plays the role of the velocity of propagation of the wave.  
This led to the conclusion that the light was an EM wave which
propagates with velocity $c$ with respect to a supporting medium and that Maxwell equations were valid in a frame connected 
to that medium. Moreover as pointed out by 
Poincar\'e and
Lorentz,  Maxwell equations are not invariant in form (\emph{covariant}) under Galilean transformations which at that time
were believed to connect inertial observers.
This would mean that the Galilean principle of relativity that Physics laws are the same for all inertial observers
would hold good only for Mechanics laws.

In his paper \cite{AE05} Einstein proposed a different solution which proved to be the correct one.

\chapter{Galilean Transformations and Classical Mechanics}
\begin{figure}[htb]
\centering
\rotatebox{0}{
\includegraphics*[width=56mm]{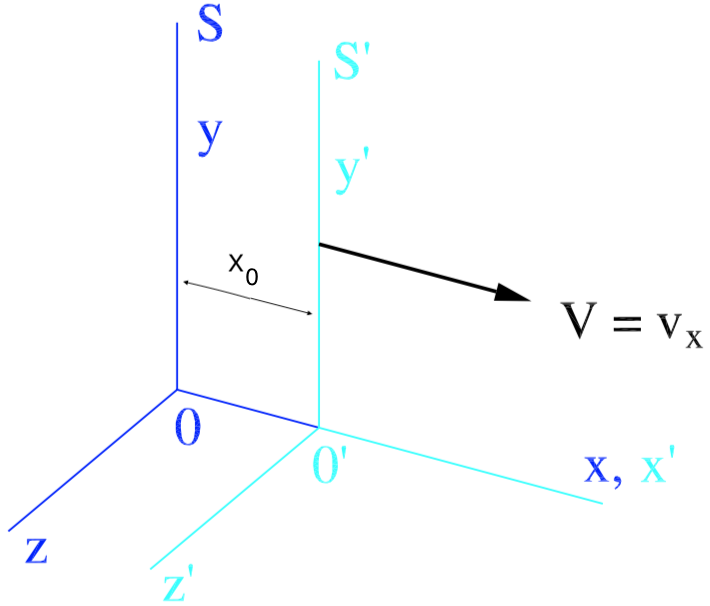}}
\caption{\label{eliana-rel-f1} The accented frame $S'$ is shifted by $x_0$ with respect to $S$.}
\end{figure}

The quantitative description of physical phenomena needs a reference frame where the coordinates of the
observed objects are specified, a ruler for measuring the distances and a clock for describing the coordinates variation with time. 
Geometry says how coordinates in two different reference frames
are related. If we assume for sake of simplicity two reference frames simply shifted along one of the
axis\footnote{All other cases can be obtained by introducing a rotation of the axis and a shift of the origin.}, 
for instance by $x_0$ along $x$, 
the relationships are  (see Fig.~\ref{eliana-rel-f1})
$$ x'=x-x_0 \hspace*{6mm} y'=y  \hspace*{6mm} z'=z $$

If $S'$ is moving along the common $x$-axis with speed $\vec V$=$\hat x V$ 
with respect to $S$, assuming the origins coincide at $t$=0  it is

\begin{eqnarray}\label{eliana-rel-eq1}
x'=x-x_0=x-Vt \hspace*{7mm} y'=y  \hspace*{6mm} z'=z  
\end{eqnarray}

\begin{figure}[htb]
\centering
\rotatebox{0}{
\includegraphics*[width=46mm,height=26mm]{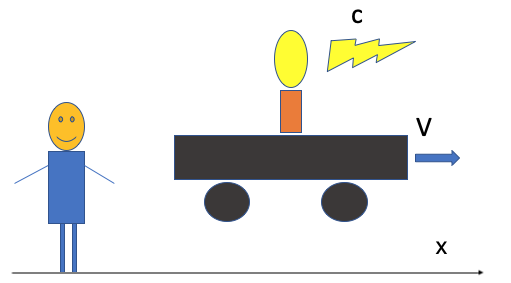}}
\caption{\label{eliana-rel-f2} Light source on a train moving along the $x$-direction with uniform speed $\hat x V$ wrt the railway
platform.}
\end{figure}

Eqs.(\ref{eliana-rel-eq1}) are the Galilean coordinate transformations. By differentiating with respect to time it is

\begin{eqnarray}\label{eliana-rel-eq2}
\dot x'=\dot x -V \hspace*{7mm} \dot y'=\dot y  \hspace*{6mm} \dot z'= \dot z  
\end{eqnarray}

where we have implicitly assumed that 
$t'$=$t$ and that the lengths are the same.
From Eqs.(\ref{eliana-rel-eq2}) we see that velocities add. 
If the light from a source on a train propagates in the $x$-direction with velocity $\hat x c$, for an observer 
at rest on the railway platform it
would propagate with velocity $\hat x (c+V)$ (see Fig.~\ref{eliana-rel-f2}).

By differentiating Eqs.(\ref{eliana-rel-eq2}) wrt time we get
\begin{eqnarray}\label{eliana-rel-eq3}
\ddot x'=\ddot x \hspace*{7mm} \ddot y'=\ddot y  \hspace*{6mm} \ddot z'= \ddot z 
\end{eqnarray}

that is the acceleration of a body is the same for all observers related by Galilean transformations.

The basic laws of classical dynamics are
\begin{enumerate}
\item A \emph{free} body perseveres in its state of rest, or of uniform motion (principle of inertia). Reference 
frame where the principle of inertia
holds good are said inertial.
\item In an inertial reference frame it is $\vec F=m\vec a$, that is the acceleration, $\vec a$, is proportional to the
applied force, $\vec F$, through a
constant, $m$ (``inertial mass''). In other words, if in an inertial frame a body appears to be accelerated it means that
there must be something acting on it.
Implicitly it is
assumed that $m$ is a characteristic of the body which
doesn't depend upon its status of motion.
\item Whenever
  two bodies interact they apply equal and opposite forces to each other.
\end{enumerate}
The second and third laws combined give the total momentum conservation for an isolated system.
The three laws of dynamics hold good in inertial frames.  
If an inertial frame exists, all reference frames in uniform motion with respect to it are inertial.
As they are all equivalent it is reasonable to assume
that all mechanics laws are the same for inertial observers (principle of relativity). More precisely, the
principle states that the laws must have the \emph{same form} (covariance). If we chose a non inertial frame for describing
the motion of an object, the numerical results would be the same if the motion of the reference frame itself is accounted for
correctly.
However the equation of motion for the observed object would take a different form.

\begin{figure}[htb]
\centering
\rotatebox{0}{
\includegraphics*[width=46mm,height=26mm]{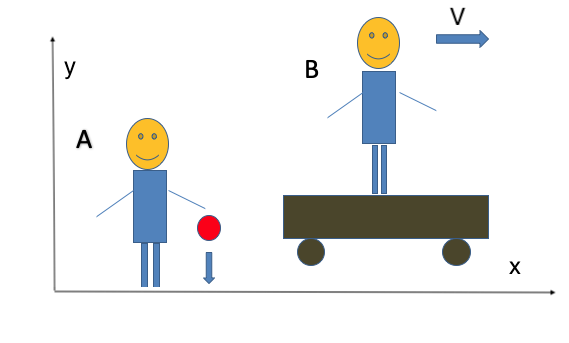}}
\caption{\label{eliana-rel-f3} Alex, at rest on the railway platform, studies the motion of objects under the gravitational force. 
Betty is on a train 
moving along the $x$-direction with uniform speed $\hat x V$ wrt the railway
platform.}
\end{figure}

Are mechanics laws invariant under Galilean transformations? Suppose that Alex is studying the 
motion of a ball  let to fall under the earth gravitational force.
Alex measures that the object is subject to a constant acceleration of $a\approx$ 9.8 ms$^{-2}$. By using different balls
he finds that the acceleration is always the same, $g$. He
concludes that there must be a force acting on the balls which is directed towards the center of the earth 
and having magnitude $mg$. Betty is on a train moving uniformly with velocity $\vec V$=$\hat x V$ with respect to Alex
(see Fig. \ref{eliana-rel-f3}).

From Eqs.(\ref{eliana-rel-eq3})

\begin{alignat}{3}
 \ddot x' & = \ddot x =0 & \qquad \ddot y' & = \ddot y \nonumber 
\end{alignat}

and as the mass, $m$ is a constant, she will agree with Alex on magnitude and direction 
of the force. 
Classical mechanics laws are covariant under Galilean transformations.


We want to show in a more formal way that Newton law $\vec F$=$m\vec a$ is invariant under Galilean transformations by using the example
of a system of particles which internal forces depend upon the reciprocal distances, 
$r_{ij}$.
In the inertial reference frame $S$ it is
$$\vec F_i = -\nabla_{r_i} \Sigma_j U(r_{ij})=m_i\vec a_i$$
In the moving frame $S'$ the Newton law must take the same form
with the potential $U$ having the same functional dependence upon the new variables as in the old ones.
From the Galilean transformations Eqs.(\ref{eliana-rel-eq1}) and (\ref{eliana-rel-eq3}) it is
$r'_{ij}=r_{ij}$,
$\vec a'_i=\vec a_i$ and 
$\nabla_{r'_i}=\nabla_{r_i}$
and therefore, as the mass is a scalar invariant,  it is indeed 
$$\vec F'_i = -\nabla_{r'_i} \Sigma_j U(r'_{ij})=m_i\vec a'_i$$

\chapter{Relativistic Kinematics}
\section{Galilean relativity and EM wave equation}
Using the cyclic rule\footnote{ 
$
{\partial \over \partial x_i}=
\sum_j{\partial x'_j\over \partial x_i}
{\partial \over \partial x'_j}
$} 
the wave equation\footnote{For simplicity we have chosen the $x$-axis along the direction of propagation.}
$$
\left[ \frac{\partial^2}{\partial x^2} - \frac{1}{c^2}\frac{\partial^2}{\partial t^2}\right]\Phi  = 0 
$$
becomes under Galilean transformation 
\begin{align*}
\left[\frac{\partial^2}{\partial x'^2} - \frac{1}{c^2} \frac{\partial^2}{\partial t'^2}
                                       - \frac{V^2}{c^2} \frac{\partial^2}{\partial x'^2}
                     - 2\frac{V}{c^2} \frac{\partial^2}{\partial x'\partial t'}\right]\Phi & = 0
\end{align*}
and it is clearly not covariant. 
As anticipated,  Maxwell equations would describe EM laws in a particular reference frame, and as such, a \emph{privileged} one.
It was conjectured the existence of a medium, the \emph{luminiferous aether}, supporting the propagation of EM waves, as the air supports sound waves.
This medium had to be extremely rarefied to be undetectable directly and it would permeate the whole space. 
The speed of light would be $c$ with respect to the medium and, accordingly to Eqs.(\ref{eliana-rel-eq2}), 
would be different for an observer 
moving with respect to the medium. 

Experiments for demonstrating the existence of the aether, by measuring the speed of light under 
different conditions were attempted, the most famous of them being
those performed by Michelson and Morley using an interferometer.
  
\begin{figure}[htb]
\centering
\rotatebox{0}{
\includegraphics*[width=46mm,height=30mm]{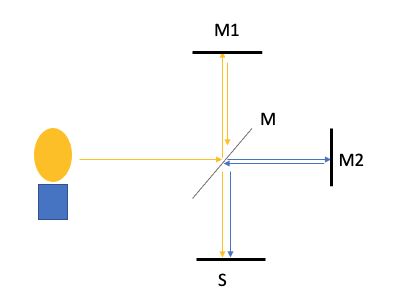}}
\caption{\label{eliana-rel-f4} A schematic view of Michelson-Morley interferometer experiment.}
\end{figure}

The arrangement is schematically shown in Fig.\ref{eliana-rel-f4}.
The light is split into two orthogonal patterns of equal 
length by the partially silvered glass M, reflected back by mirrors M1 and M2 and recombined on a screen.
If the earth is at rest in the aether, the recombined waves are in 
phase but if the earth is moving the time needed by the two waves for reaching the 
screen  would be different and an interference pattern should be observed on the screen S.
While rotating around the sun, the earth motion direction changes and it should be possible to observe interference patterns,
at least in some periods of the year.

For avoiding errors due to incorrect mirrors angle or to the distances between the two 
mirrors and the partially silvered glass being not identical, the apparatus can
be rotated so that the possible interference fringes would move.
 
The result of the first experiment in 1887 was negative. 
It was repeated with higher accuracy apparatuses during the following 50 years, however
the result was always negative. Theories proposed to justify the negative result were contradicted by other experiments. 
A detailed quantitative description of these experiments may be found in \cite{RR68} 

Attempts of modifying  the still relatively new EM laws in such a way that they
would be invariant under Galilean transformations
led to predictions of new phenomena which could not be proved experimentally.

\section{Einstein Postulates}

In 1905 Einstein\cite{AE05}  proposed a solution to the dilemma based on two postulates:

\begin{enumerate}
\item Physics laws are the same in all inertial frames, there is no
          preferred reference frame.
\item The speed of light in the empty space has the same finite value
          $c$ in all inertial frames.
\end{enumerate}

At that time the existence of the aether was still widely accepted and not yet ruled out by experiments.
It is worth noting that Lorentz had found the coordinates transformation which leave Maxwell's equations invariant in 1904,
before the publication of Einstein's paper, accompanied however by an erroneous interpretation.
It is in  Einstein paper that such transformations are \emph{physically} justified and therefore 
extendable to the whole Physics. In particular,
the concept of time was critically addressed and the fact that the time is not universal comes as a consequence of the
light having a finite velocity. 

\begin{figure}[htb]
\centering
\rotatebox{0}{
\includegraphics*[width=54mm,height=34mm]{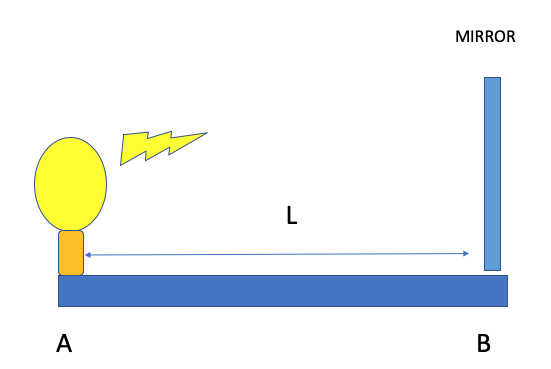}}
\caption{\label{eliana-rel-f5} Synchronization procedure of the clocks in $A$ and $B$.}
\end{figure}

Let us summarize Einstein reasoning.
In order to describe the motion of an object we need to equip each point of our reference frame with identical  clocks and rulers.
Is it possible to synchronize the clocks by sending light rays. For instance we can imagine of sending a light ray from a point $A$ to
$B$ and $B$ reflecting it back to $A$ (see Fig.\ref{eliana-rel-f5}). The two observers sitting in $A$ and $B$ may agree in 
setting the clock in $B$ at the arrival of the signal to a given value
$t_B$ while $A$ will set its own clock to 2$t_B$ when receiving back the signal. However if we want the speed of light 
to be $c$=3$\times$10$^8$
m s$^{-1}$ we shall measure the distance, $L$,  between $A$ and $B$ and set $t_B$=$L/c$.

Assuming the clocks are identical, they will stay synchronized. For this procedure we use the light because, we have assumed that
it propagates in vacuum with constant velocity so that we can be assured that the velocity is the same in both directions.

Once all clocks within one frame are synchronized we can establish the chronological sequence between 
events happenings in different places within the same frame of reference.

\begin{figure}[htb]
\centering
\rotatebox{0}{
\includegraphics*[width=54mm,height=34mm]{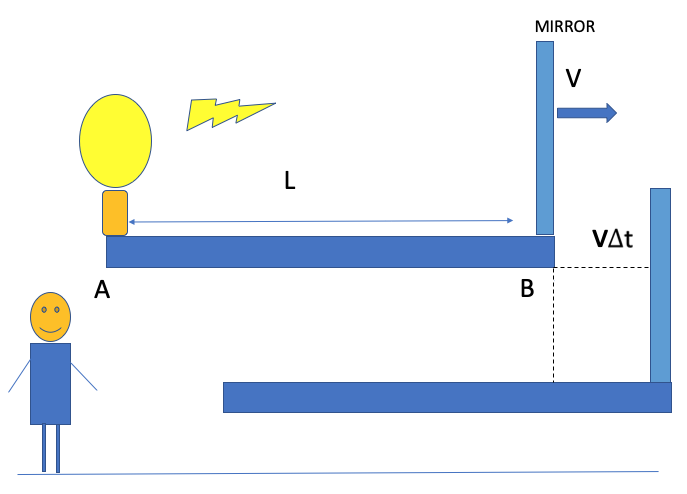}}
\caption{\label{eliana-rel-f6} Synchronization procedure of $S'$ clocks as seen by the ``resting" observer.}
\end{figure}

The observer $S'$ moving with respect to $S$ may synchronize its own clocks with the very same procedure.
However this synchronization procedure observed by the resting observer is not correct. Suppose $A$ and $B$ lying on
the common $x$-axis with $B$ on the right of $A$ ($x_B>x_A$) as shown in Fig.\ref{eliana-rel-f6}: while the light moves to $B$, $B$ moves further away
and once reflected back to $A$, $A$ moves toward the light. Therefore observed by $S$ the time needed to reach $B$ is
obtained by setting
$$c t_B=L+V t_B $$
($L\equiv x_B-x_A$) which gives
$$ t_B=L/(c-V)$$
while the time needed to reach $A$  is obtained from
$$c t_A=L-V t_A$$
that is
$$t_A=L/(c+V)$$
and
$$ t_B-t_A=L\Bigl(\frac{1}{c-V}-\frac{1}{c+V}\Bigr)=\frac{2VL}{c^2[1-(V/c)^2]}\neq 0 $$

Therefore for $S$, $S'$ clocks are not synchronized. 
If the clocks in the moving frame would be synchronous with the stationary ones
they wouldn't be synchronous in their own frame. The ``stationary" frame would dictate the timing.
However stationarity is relative, the inertial frames are all equivalent: if there exist no privileged 
frame, we must abandon the idea of universal time. Relativity of time is a consequence of the speed of light being finite.

As a consequence events which may be simultaneous for $S$ are in general not
simultaneous for $S'$ and the
other way round.

\section{ Lorentz transformations}
By assuming the speed of light constant in all reference
frames, the Galilean
transformations, implying the addition of velocity rule, must
be modified. The new transformations must reduce to the Galilean ones
when the relative motion is slow ($V\ll c$). 
According to the first Einstein postulate,
the empty space is isotrope (all direction
are equivalent) and homogeneous (all points are equivalent);
it would make no sense to postulate that the laws are
invariant in a space which is not homogeneous and isotrope.
As time is not universal, it must be included in the coordinate transformation.

Resorting to arguments of space homogeneity and isotropy, and  to the Einstein postulates
it is relatively simple to work out the correct coordinates transformation.

Homogeneity implies the relationship between the coordinates must be linear:
\begin{align*}
x' & = a_{11} x + a_{12} y + a_{13} z + a_{14} t \\
y' & = a_{21} x + a_{22} y + a_{23} z + a_{24} t \\
z' & = a_{31} x + a_{32} y + a_{33} z + a_{34} t \\
t' & = a_{41} x + a_{42} y + a_{43} z + a_{44} t
\end{align*}
where the coefficients $a_{ij}$ may depend upon the relative speed $V$.

The points on the $x$-axis where $y$=$z$=0 must transform to $y'$=$z'$=0 at all times which means that 
$a_{21}$=$a_{31}$=$a_{24}$=$a_{34}$=0. The points with $y$=0 (the $x$-$z$ plan) must transform into $y'$=0
and therefore  it is also $a_{23}$=0. The points with $z$=0 (the $x$-$y$ plan) must transform into $z'$=0
and therefore  it is also $a_{32}$=0. Because of  isotropy,
time must be invariant for a 
sign inversion of the coordinates  $y$ and $z$ which means $a_{42}$=$a_{43}$=0.
So we are left with 8 unknown coefficients:
\begin{align*}
x' & = a_{11} x + a_{12} y + a_{13} z + a_{14} t \\
y' & = a_{22} y  \\
z' & = a_{33} z  \\
t' & = a_{41} x + a_{44} t
\end{align*}
For a point on the $y$-axis ($x$=$z$=0) it is
$$x'=a_{12}y+a_{14}t
$$
and therefore $x'$ value would depends on the sign of $y$ which again contradicts the hypothesis of isotropy. 
Therefore it must be $a_{12}$=0. The same argument can be used to set $a_{13}$=0.
We are left with
\begin{align*}
x' & = a_{11} x + a_{14} t \\
y' & = a_{22} y \\
z' & = a_{33} z \\
t' & = a_{41} x + a_{44} t
\end{align*}
The value of $a_{22}$  is found by observing that
$$y'=a_{22}(V)y=a_{22}(V)a_{22}(-V)y'$$
that is $a_{22}(V)a_{22}(-V)$=1. Because $a_{22}$=1 for $V\rightarrow 0$, the correct choice is 
$a_{22}$=1.  In the same way it is found $a_{33}$=1.

The origin of the $S'$ frame is described in $S$ as $x=Vt$ and has by definition $x'$=0  at any time. 
Therefore
$$0=x'_0=a_{11}x_0+a_{14}t=a_{11}Vt+a_{14} t $$
that is $a_{14}$ and $a_{11}$ are related by
$$a_{14}/a_{11}=-V$$
and the equation for $x'$ becomes
$$x'=a_{11}(x+a_{14}t/a_{11})=a_{11}(x-Vt)$$

For finding the values of the remaining coefficients  $a_{11}, a_{41}$ and  $a_{44}$
we resort to the fact that the speed of light is the same in $S$ and $S'$
and that the wave equation is invariant in form. Suppose an EM spherical wave leaves the origin of the frame 
$S$  at $t$ = 0.
The propagation is described in $S$ by the equation of a sphere which radius squared increases with time as
\begin{equation}\label{eliana-rel-eq4}
R^2(t)=x^2+y^2+z^2=c^2t^2
\end{equation}
In $S'$ the wave propagates with the same speed $c$ and therefore
$$R'^2(t')=x'^2+y'^2+z'^2=c^2t'^2$$
which writing the primed coordinates $x'$, $y'$, $z'$ and $t'$ in terms of the un-primed ones becomes

$$a_{11}^2x^2 +a_{11}^2V^2t^2 - 2a_{11} xVt+y^2+z^2=c^2a_{41}^2 x^2+ c^2a_{44}^2t^2 + 2a_{41}a_{44}xt$$
Rearranging the terms it is
$$(a_{11}^2 - c^2a_{41}^2)x^2  - 2(a_{11}^2V+c^2a_{41}a_{44})xt+y^2+z^2= (c^2a_{44}^2-a_{11}^2V^2)t^2 $$
Comparing this equation with Eq.(\ref{eliana-rel-eq4}), we get a system of 3 equations in the 3 unknown $a_{11}$, $a_{41}$ and $a_{44}$

\begin{align*}
a_{11}^2 - c^2a_{41}^2 & = 1 \\
a_{11}^2V+c^2a_{41}a_{44} & =  0 \\
c^2a_{44}^2-a_{11}^2V^2 & = c^2
\end{align*}

which is solved by
$$a_{11}=a_{44}=\frac{1}{\sqrt{1-(V/c)^2}}$$
$$a_{41}= - \frac{V/c^2}{\sqrt{1-(V/c)^2}}   $$


The final coordinates transformations for a uniform motion along the common $x$-axis with relative speed $V$ are therefore
(Lorentz transformations)

\begin{equation}\label{eliana-rel-eq5}
x' =\gamma(x - \beta ct)  \hskip 1 cm y'=y \hskip 1 cm z'=z \hskip 1 cm  ct' = \gamma (ct - \beta x)
\end{equation}
with
$$\beta\equiv V/c  \hspace*{8mm} \mbox{and}  \hspace*{5mm} \gamma\equiv\frac{1}{\sqrt{1-\beta^2}} $$

The inverse transformation from $S'$ to $S$ is obtained by replacing $\beta$ with $-\beta$.

\begin{figure}[htb]
\centering
\rotatebox{0}{
\includegraphics*[width=76mm]{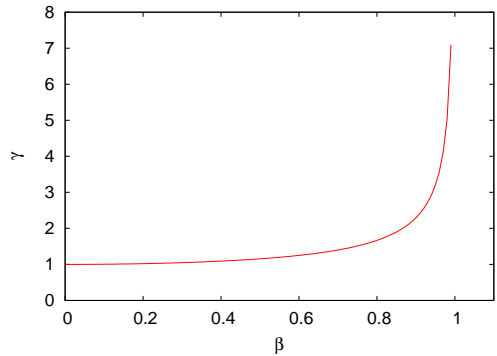}}
\caption{\label{eliana-rel-f7} $\gamma$ as function of $\beta\equiv v/c$.}
\end{figure}

It is worth noting that for $V\ll c$, that is $\beta\rightarrow$0 and $\gamma\rightarrow$1, 
Lorentz transformations coincide with the Galilean ones,
while if $V>c$, $\gamma$ becomes imaginary and the transformations  are meaningless. 
Therefore $\beta$ and $\gamma$ range between 0 and 1 and 1 and $+\infty$ respectively.
The fact that $c$ is the limit velocity
is not an Einstein postulate, it is a consequence of the Lorentz transformation.  
Fig.~\ref{eliana-rel-f7} shows $\gamma$ as function of $\beta$.

The general expression of the Lorentz transformation of parallel translation with arbitrary direction 
of the relative velocity  reads~\cite{DJ75}

\begin{equation}\label{eliana-rel-eq5a}
ct'  = \gamma \bigl(ct - \vec{\beta} \cdot \vec{r}\bigr)
\hspace*{14mm}
\vec{r}\hspace*{1mm}'  = \vec{r} +\frac{\gamma-1}{\beta^2}
 \vec{\beta} \cdot \vec{r} \vec{\beta}-\gamma \vec{\beta} c t
\end{equation}
with $\vec{\beta} \equiv {\vec{V}}/{c}$.

In matrix form Eq.(\ref{eliana-rel-eq5}) writes
$$ \left (
\begin{matrix} ct' \\
                       x' \\
                       y' \\
                       z'  \\
                        \end{matrix}
                         \right )=\gamma \left (
\begin{matrix} 1 &  -\beta & 0 & 0 \\
                       -\beta & 1 & 0 & 0 \\
                       0 & 0 & 1 & 0 \\
                       0 & 0 & 0 & 1   \\
                                  \end{matrix}
                                  \right )
 \left (  
 \begin{matrix} ct \\
                        x \\
                       y \\
                       z \\
                        \end{matrix}
                         \right )  
                         \equiv
                          {\cal L}     \left (  
\begin{matrix} ct \\                                
                       x \\
                       y \\
                       z \\
                        \end{matrix}
                         \right )                                                                                                 
                                  $$
                              
Successive Lorentz transformations may be obtained by matrix multiplication.
Let's consider a frame $S'$ moving with velocity  $\hat x V_1$ with respect to $S$ and a third frame, $S''$, 
moving with velocity $\hat x' V_2$ with respect to $S'$ and $\hat x V_3$ with respect to $S$.
The transformation from $S$ to $S''$ may be written as
 
$$ 
 {\cal L}_{S\rightarrow S''}=  
\gamma_3 \left (
\begin{matrix} 1 &  -\beta_3 & 0 & 0 \\
                       -\beta_3 & 1 & 0 & 0 \\
                       0 & 0 & 1 & 0 \\
                       0 & 0 & 0 & 1   \\
                                  \end{matrix}
                                  \right )
  =                                
 \gamma_1\gamma_2 \left (  
 \begin{matrix} 1 +\beta_1\beta_2& -\beta_1 -\beta_2 & 0 & 0 \\
                       -\beta_1-\beta_2 & 1 +\beta_1\beta_2& 0 & 0 \\
                      0 & 0 & 1 & 0 \\
                      0 & 0 & 0 & 1   \\
                                 \end{matrix}
                                  \right )                                                                                 
                                  $$
It is easy to verify that  $V_3$=$V_1+V_2$  only if $ V_1 V_2 << c^2$. Relativistically
velocities do not add.

Time is one of the 4 coordinates describing an event and as the spatial coordinates is subject to
a (Lorentz) transformation between moving frames. 

For spatial coordinates it is always possible  if for instance  
$x_2 > x_1$ to find a new coordinates frame such that
$x'_2 < x'_1$.

Is it possible to find 
a Lorentz transformation which inverts the temporal order of events?

Assume an event happening at the time $t_1$ at the location $x_1$ in $S$
and  a second event happens at $t_2$ in $x_2$ with $t_2>t_1$. 
Is it possible to find a Lorentz transformation
such that $t'_2<t'_1$? In $S'$ it is
$$ct'_{1}=\gamma(ct_{1}-\beta x_{1}) $$
$$ct'_{2}=\gamma(ct_{2}-\beta x_{2} ) $$
and therefore
$$c(t'_2-t'_1)=\gamma[c(t_2-t_1) -\beta(x_2-x_1)]$$
Therefore it is $t'_2<t'_1$  if $\beta(x_2-x_1)>c(t_2-t_1)$,  that is if $V(x_2-x_1)/(t_2-t_1)>c^2$.
This may be possible depending on the values of $x_2-x_1$ and $t_2-t_1$. 
However if the first event in $S$ \emph{drives} the second one,
$x_2$ and $t_2$ are not arbitrary.
\newline If $w$ is the speed of the signal triggering the second event from the first one it is
$$x_2-x_1=w(t_2-t_1)$$ 
\vspace*{-8mm}
$$c(t'_2-t'_1)=\gamma[c(t_2-t_1) -\beta w (t_2-t_1)]=\gamma c (t_2-t_1) \Bigl(1-\frac{V w}{c^2}\Bigr)
$$
which is always positive as $w\leq c$. Causality is not violated.

\section{Some consequences of Lorentz transformations:\\ length contraction and time dilation}
As a consequence of Lorentz transformations,
lengths are not invariant. Consider for instance a rod along the $x$-axis and at rest in the moving frame $S'$. The length of the rod in $S'$ is $L'$.
The length in $S$ is determined by the positions of the rod ends at the \emph{same} time and therefore from Eq.(\ref{eliana-rel-eq5})
with $t_1$=$t_2$

$$L'=x'_2-x'_1= \gamma(x_2-x_1)=\gamma L \hspace*{4mm} \rightarrow  \hspace*{2mm} L=L'/\gamma $$

The moving rod is shorter  than in the frame where it is at rest (\emph{length contraction}). 
However the length of a rod aligned with one of the two axis perpendicular to the direction of motion is invariant.
For this reason angles are in general not invariant.

Suppose a clock at rest in $S$ measuring a time interval $t_2-t_1$ between two events
happening at that same location in $S$. From Eq.(\ref{eliana-rel-eq5}) with $x_1$=$x_2$ the time interval in $S'$ between the 
two events is
$$t'_2-t'_1=\gamma (t_2-t_1)$$
which is larger than measured in $S$ (\emph{dilation of time}).
Moreover events happening at the same
time but in \emph{different places} in $S$, will be no more simultaneous in the moving frame $S'$. In fact using Eq.(\ref{eliana-rel-eq5})
with $t_1$=$t_2$  it is

$$c(t'_2-t'_1)=\gamma\beta (x_1-x_2)$$
which is non vanishing unless $x_1$=$x_2$.

In general it is named as \emph{proper} the interval (in space or time) measured in a inertial frame where the observed object is at
rest. 

\vspace*{4mm}
Let us suppose that we have two synchronized clocks, $C1$ and $C2$, at the origin $O$ of $S$ and that at $t$=0 we set 
$C2$ in uniform 
motion along the $x$-axis with velocity $V$. 
After a time $t_{C1}$=$T$, when $C2$ accordingly to time dilation strikes $T/\gamma$, the clock $C2$ inverts its direction. 
When $C2$ arrives back in $O$,  $C1$ strikes
$2T$ and $C2$ instead 2$T/\gamma$.  
This may look as a paradox because the notion of motion is relative: with respect to $C2$, 
it was $C1$ moving 
and therefore $C2$ should strike 2$T$ and $C1$ instead 2$T/\gamma$. However when they are both in $O$ we can compare 
their time and only one outcome is possible. 
The mistake is considering the two situations to be equivalent, while they are not.
$C2$ has been set in motion by the action of some kind of force and some kind of force also is responsible for changing 
its direction,
while $C1$ has experienced no force. Indeed direct experiments involving clocks have shown that time dilation is real~\cite{HK72}.

\section{Lorentz transformations for velocity and acceleration}
The relativistic transformation for the velocity follow from
the Lorentz transformations for the coordinates
$$
v'_x  \equiv \frac{dx'}{dt'}  = \frac{dx-Vdt}{dt-Vdx/c^2}
          = \frac{v_x-V}{1-v_x\beta/c} 
          $$
\begin{equation}\label{eliana-rel-eq5b}          
v'_y \equiv \frac{dy'}{dt'}  = \frac{dy}{\gamma(dt-Vdx/c^2)}
          = \frac{v_y}{\gamma (1-v_x \beta/c)} 
         \end{equation}
$$         
v'_z \equiv \frac{dz'}{dt'}  = \frac{dz}{\gamma(dt-Vdx/c^2)}
          = \frac{v_z}{\gamma (1-v_x \beta/c)}
         $$

\noindent
with $\beta\equiv V/c$, $v_x\equiv dx/dt$, $v_y\equiv dy/dt$ and $v_z\equiv dz/dt$. 
The inverse transformation is obtained by replacing $V$ with $-V$.
\noindent
Unlike the classical case,
also the components of the velocity perpendicular to
the motion, when non vanishing,
are affected by the motion.
This is a consequence of the
fact that the time is not invariant and therefore, although the lengths
perpendicular to the motion direction are unchanged, the time needed
to cover them is changed.

As an exercise, let us use these expressions for a light ray. 
For $v_x$=$c$ and $v_y$=$v_z$=0 it is 
$$v'_x=\frac{c-V}{1-V/c}=c\hspace*{1mm}\frac{c-V}{c-V}=c \hspace*{4mm} \mbox{and } \hspace*{2mm} v'_y=v'_z=0 $$
For $v_y$=$c$ and $v_x$=$v_z$=0 it is $v'_x$=$-V$, $v'_y$=$c/\gamma$, $v'_z$=0 and
$$v_x'^2+v_y'^2+v'^2_z=V^2+c^2[1-(V/c)^2]= c^2$$
As expected, the speed of light is invariant.

In a similar way as for the velocity it is possible to find
the transformation for the acceleration~\cite{RR68}

$$a'_x=\frac{a_x}{\gamma^3(1-v_x\beta/c)^3 }$$
\begin{equation}\label{eliana-rel-eq6}
a'_y=\frac{a_y}{\gamma^2(1-v_x\beta/c)^2 } + \frac{a_x v_y\beta/c}{\gamma^2(1-v_x\beta/c)^3 }
\end{equation}
$$a'_z=\frac{a_z}{\gamma^2(1-v_x\beta/c)^2 } + \frac{a_x v_z\beta/c}{\gamma^2(1-v_x\beta/c)^3 }$$

 Acceleration is not invariant under Lorentz transformations
 unless both $v$ and $V$ $\rightarrow$ 0.

\section{Experimental evidence of relativistic kinematics}
In his papers Einstein suggested  possible experiments for confirming the validity of his theory. Here we give some examples:
light aberration, (transverse) Doppler effect and lifetime of unstable particles.

\subsection{Light aberration}
\begin{figure}[htb]
\centering
\rotatebox{0}{
\includegraphics*[width=78mm]{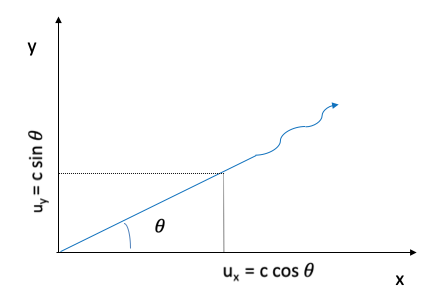}}
\caption{\label{eliana-rel-f8} Source emitting a light ray at an angle $\theta$ with respect to the $x$-axis in $S$.}
\end{figure}

Light aberration is the apparent motion of a light source due to the movement of the observer. It was first
discovered in astronomy. 
Consider a source emitting photons at an angle $\theta$ with respect to the $x$-axis in the $S$ frame where
$v_y$=$c\sin{\theta}$ and $v_x$=$c\cos{\theta}$ (see Fig.\ref{eliana-rel-f8}).
In $S'$ it is $v'_y$=$c'\sin{\theta'}$ and $v_x$=$c'\cos{\theta'}$.
Using Galilean
transformations for the velocity components 

$$v'_y=v_y  \hspace*{2mm} \mbox{and} \hspace*{2mm} v'_x=v_x-V$$
it is
$$\tan{\theta'}=v'_y/v'_x=v_y/(v_x-V)$$
and
$$\tan{\theta'}=\frac{\sin{\theta}}{(\cos{\theta}-\beta)}$$
Using instead Lorentz transformations Eq.(\ref{eliana-rel-eq5b})
$$\tan{\theta'}=\frac{\sin{\theta} }{\gamma(\cos{\theta}-\beta)}$$

\begin{figure}[htb]
\begin{minipage}[c]{0.4\linewidth}
\rotatebox{0}{
\includegraphics*[width=78mm]{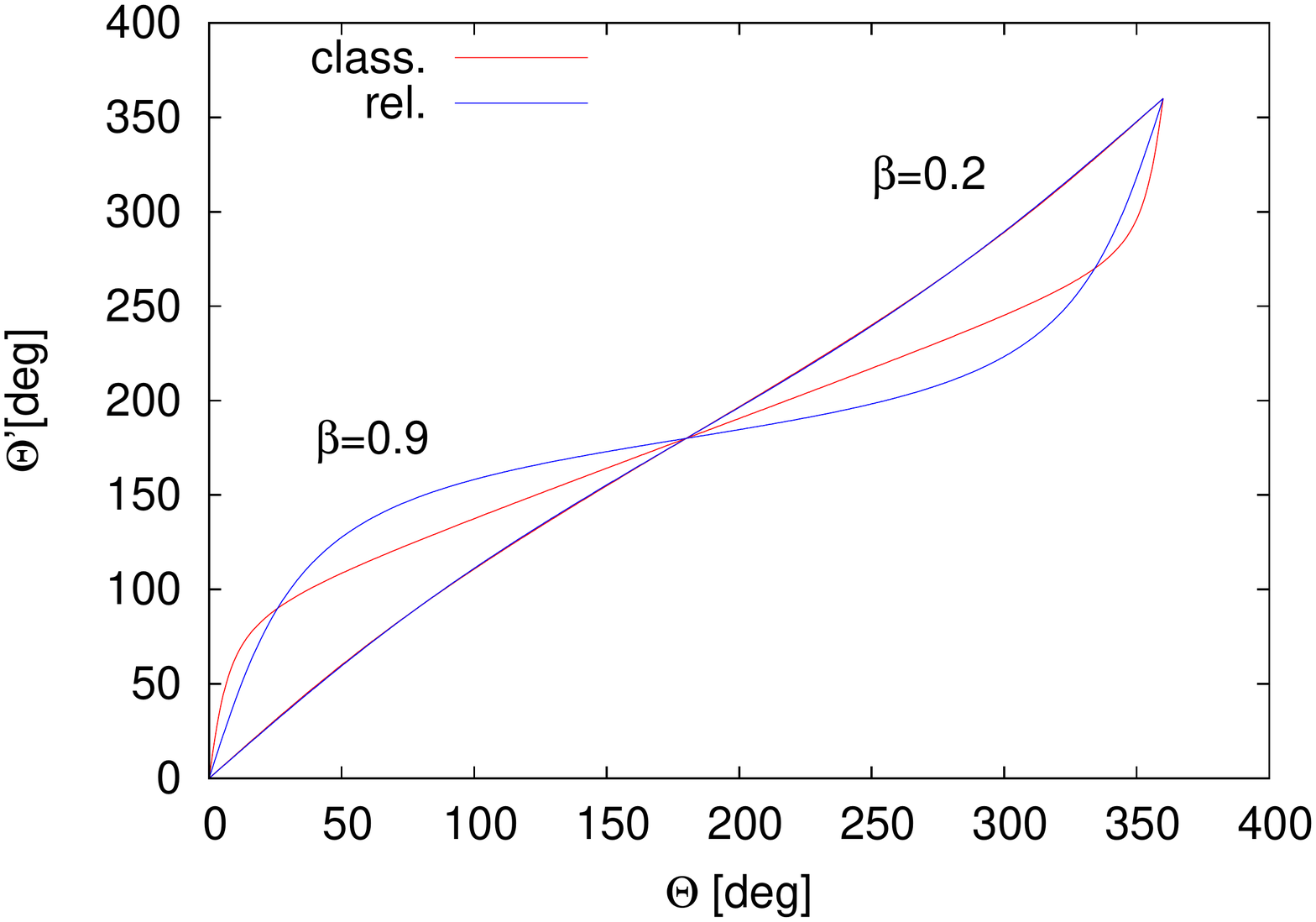}}
\caption{\label{eliana-rel-f9} Angles observed in the moving frame for $\beta$=0.2 and $\beta$=0.9.} 
\end{minipage}
\hfill
\begin{minipage}[c]{0.4\linewidth}
\rotatebox{0}{
\includegraphics*[width=78mm]{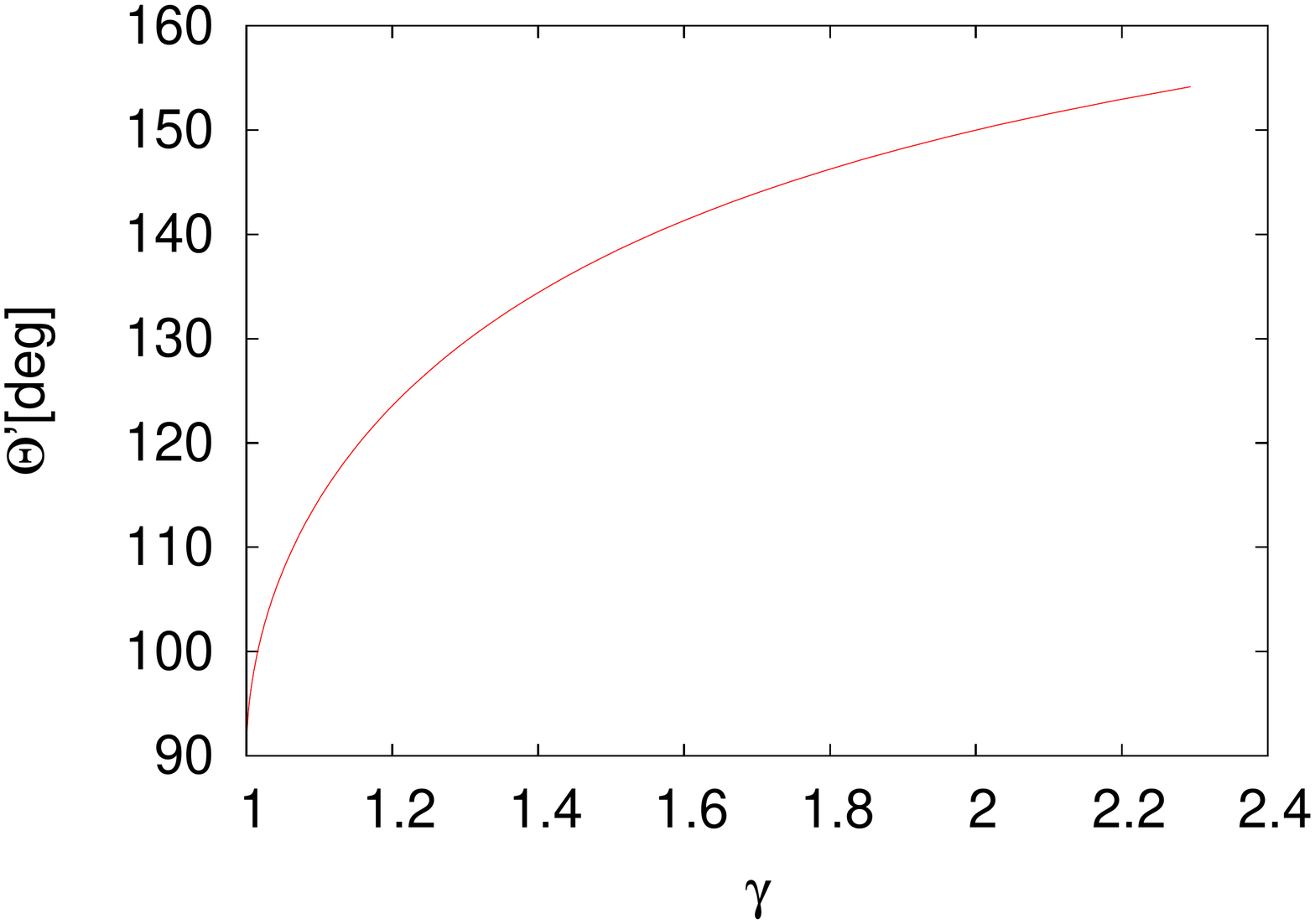}}
\caption{\label{eliana-rel-f9a} $\theta'$ vs. $\gamma$ for $\theta$=90$^0$. }
\end{minipage}%
\end{figure}

High energy experiments involving emission of photons confirm the relativistic expression.
Fig.\ref{eliana-rel-f9} shows the classical and relativistic relationships between 
the emission angles  for $\beta$=0.2
and $\beta$=0.9. 
We see that $\theta'$=$\theta$
for 0 and 180 degrees for both the classical as well as the relativistic expression.
In all other cases it must be paid attention whether the angles are specified in the moving or in the rest frame. 
Fig.\ref{eliana-rel-f9a}  shows how the angle $\theta$=90$^0$ transforms as function of $\gamma$.

\subsection{Doppler effect}

In the following we give an alternative computation of light aberration  by using the undulatory description
of light which allows also to treat the Doppler effect.

\noindent 
Consider a plane light wave propagating in the direction $\hat r=\hat x \cos{\theta}+\hat y \sin{\theta}$

\begin{eqnarray}\label{eliana-rel-eq8}
A(x,y;t)=\cos{[k(x \cos{\theta}+ y \sin{\theta})-\omega t ]}
\end{eqnarray}
where $k$=$\omega/c$  is the wave number. 
The wave must have the same form when observed in $S'$

$$A(x',y';t')=\cos{[k'(x' \cos{\theta'}+ y' \sin{\theta'})-\omega' t']}$$
Expressing the un-primed coordinates  in terms of the primed ones, Eq.(\ref{eliana-rel-eq8})
gives

$$B(x',y';t')=\cos{\{[k\gamma (x'+\beta ct)\cos{\theta}+ y' \sin{\theta})]-\omega \gamma(t'+\beta x'/c )\}}$$
Comparing with the expression for $A(x',y';t')$ we get
\begin{eqnarray}\label{eliana-rel-eq9}
k'  \cos{\theta'}=k\gamma \cos{\theta}-\omega\gamma\beta/c = k\gamma(\cos{\theta}-\beta) 
\end{eqnarray}
\begin{eqnarray}\label{eliana-rel-eq10}
k' \sin{\theta'}=k\sin{\theta}
\end{eqnarray}
\begin{eqnarray}\label{eliana-rel-eq11}
\omega'=-k\gamma\beta c \cos{\theta}+\gamma \omega = \gamma \omega (1-\beta\cos{\theta})
\end{eqnarray}

From Eqs.~(\ref{eliana-rel-eq9}) and (\ref{eliana-rel-eq10}) it is

$$\tan{\theta'}=\frac{\sin{\theta} }{\gamma(\cos{\theta}-\beta)}$$
which is the result found previously.  In addition   
Eq.~(\ref{eliana-rel-eq11}) gives the wavelength measured by two observers
in relative motion. Suppose that the source is at rest in $S$ so that $\omega$ is the proper frequency, $\omega_0$.
Thus it is
$$\omega'=\omega_0\gamma(1-\beta\cos{\theta})$$
where $\theta$ is the propagation angle in the source reference frame.
For $\theta$=0 it is
$$\omega'=\omega_0 \sqrt{\frac{1-\beta}{1+\beta}} $$
and therefore $\omega' < \omega_0$ for $\beta>0$ (receiver moving away from source), while $\omega' > \omega_0$
for $\beta<0$ (receiver moving towards the source).

\noindent 
For $\theta$=90$^0$ (in the frame where the source is at rest) it is 
$$\omega'=\omega_0\gamma$$
Unlike the classical case, relativistically it is expected 
the existence of a transverse Doppler effect which is a consequence of time being not invariant.
This was predicted by Einstein who also suggested in 1907 
an experiment using hydrogen ions for measuring it.
The experiment realized for the first time by Ives and Stilwell in 1938 proved the correctness of Einstein
prediction.

\subsection{Lifetime of unstable particles}

Beside $e^-$, $p$ and $n$, in nature there are particles which are produced by 
scattering process
and unlike $e^+$, $\bar p$  and $\bar n$, are ``short-living''. Their number decays in time as

$$N(t)=N_0 e^{-t/\tau}$$
Pions for instance are produced by bombarding a proper target by high energy protons and
leave the target with 
$v\approx$ 2.97$\times$10$^{8}$ m/s that is $\beta$=0.99 and $\gamma\approx$ 7. 
The lifetime of charged pions at rest is $\tau_0$=26$\times$10$^{-9}$ s. 
The time, $\bar t$,  needed for the pions at rest to decay by  half is
$$N(\bar t)=N_0 e^{-\bar t/\tau}=\frac{N_0}{2} \hspace*{4mm} \rightarrow \hspace*{2mm} \bar t=18 \hspace*{1mm} \mathrm{ns}$$
It is observed that they are reduced to the half after 37 m from the target.
If their lifetime would be as when at rest they should become the half 
already after about 5 m. 
\noindent
The experimental observation is explained if the pion lifetime in the laboratory frame is 
$$\tau=\gamma \tau_0$$
as predicted by  time dilation. 
\noindent
The decaying pions produce muons which are unstable too. Their  lifetime at rest is 2.2 $\mu$s, 
which is small however larger than pion one.
Time dilation may allow us  realizing future colliders smashing muons if we are able to accelerate them to high energy very quickly!

\chapter{Relativistic Dynamics}
Assuming 
$\vec F$ invariant and $m$ constant, Newton law,  $\vec F=m\vec a$,
is not invariant under Lorentz
transformations because, as we have seen, $\vec a$ is not invariant. 
\noindent
In addition the mass cannot be a constant because by applying a constant force 
to an object its speed would 
increase indefinitely becoming larger than $c$.
Classical mechanics must be modified to achieve invariance under Lorentz
transformations and the 
new expressions must reduce to the classical ones when the speed of the objects is much smaller than $c$.

\section{The relativistic mass}
In the 1905 paper, Einstein used the 
Lorentz force and the electro-magnetic field transformations to achieve 
the generalization of the definition of momentum 
and energy. 

In 1909 MIT professors of chemistry  
Lewis and Tolman~\cite{LT09} suggested a more straightforward  
reasoning with respect to Einstein original one and which involves purely mechanical  arguments. 

\begin{figure}[htb]
\centering
\rotatebox{0}{
\includegraphics*[width=72mm]{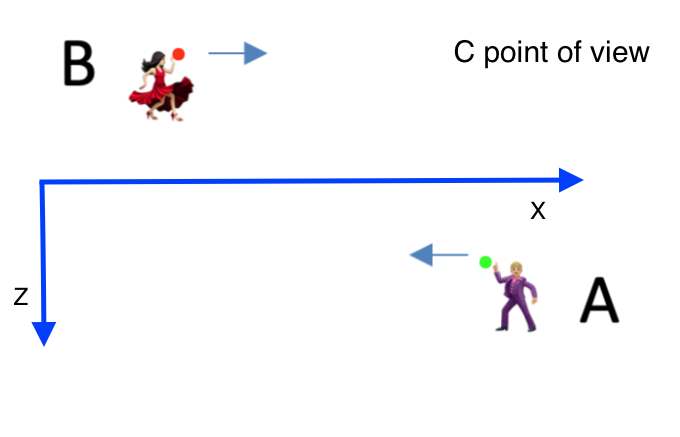}}
\caption{\label{eliana-rel-f11} Betty and Alex frames as seen by the third observer Charlie. They move
towards each other with equal and opposite velocity along the $x$-axis.}
\end{figure}

\begin{figure}
\begin{minipage}[c]{0.4\linewidth}
\rotatebox{0}{
\includegraphics*[width=42mm]{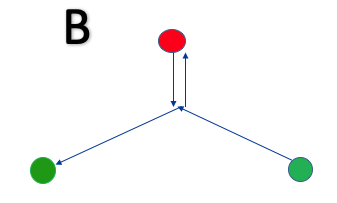}}
\caption{\label{eliana-rel-f12} The elastic collision as observed by Betty.}
\end{minipage}
\hfill
\begin{minipage}[c]{0.4\linewidth}
\rotatebox{0}{
\includegraphics*[width=42mm]{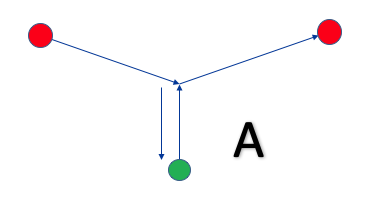}}
\caption{\label{eliana-rel-f13} The elastic collision as observed by Alex.}
\end{minipage}%
\end{figure}

\begin{figure}[htb]
\centering
\rotatebox{0}{
\includegraphics*[width=42mm]{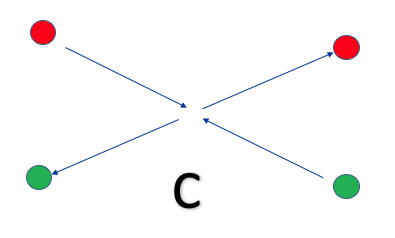}}
\caption{\label{eliana-rel-f14} The elastic collision as observed by Charlie. The two balls are scattered
under the same angle  of incidence.}
\end{figure}

Let us assume there are two observers, Alex and Betty, moving towards each other 
with the same velocity as seen by a third observer, Charlie
(see Fig.~\ref{eliana-rel-f11}).
\noindent
Betty sits in $S$ and Alex in $S'$.
Alex and Betty have identical elastic balls. 
Betty (see Fig.~\ref{eliana-rel-f12}) releases the red ball with $v^B_x$=0 and $v^B_z$=$u\neq 0$, while Alex (Fig.~\ref{eliana-rel-f13})
releases the green ball with 
speed $v'^A_x$=0 and $\bar v'^A_z$ numerically equal and opposite to the red ball velocity, that is 
before the collision it is
$$v'^A_z=-v^B_z =-u$$ 
The experiment is set so up 
that the two balls collide and rebound as shown in Fig.~\ref{eliana-rel-f14}.
\newline Now let us consider Betty point of view.
For Betty it is $\Delta v^B_x=0$ and
\begin{eqnarray}\label{eliana-rel-eq11a}
\Delta v^B_z=-2u
\end{eqnarray}

As we know the values of the velocity components of Alex ball  in the moving frame
we need here the inverse of the velocity transformation Eq.~(\ref{eliana-rel-eq5b}), that is
$$ v_x=  \frac{v'_x+V}{1+v'_x \beta/c} \hspace*{12mm}
 v_z=  \frac{v'_z}{\gamma (1+v'_x \beta/c)} $$
In our case $v'^A_x$=0 before and after the collision 
while $v'^A_z$=$-v^B_z$=$-u$ before the collision and $v'^A_z$=$u$ after.
Therefore
$v_x^A=V$ before and after the collision, and 
\begin{eqnarray}\label{eliana-rel-eq12}
v^A_z=v'^A_z/\gamma=-u/\gamma 
\end{eqnarray} 
with $ \gamma=1/\sqrt{1-(v^A_x/c)^2}$, before the collision and
\begin{eqnarray}\label{eliana-rel-eq12a}
v^A_z=u/\gamma 
\end{eqnarray} 
after.

\noindent
Momentum,
classically defined as $\vec p$=$m\vec v$, 
is conserved if $\Delta(\vec p^A + \vec p^B)$=0. In our case the $x$ component is always conserved 
as $\Delta p_x^A $=$\Delta p_x^B $=0. In addition it must be
$$ \Delta p^B_z=-\Delta p^A_z$$
which using Eqs.~(\ref{eliana-rel-eq11a}),~(\ref{eliana-rel-eq12}),~(\ref{eliana-rel-eq12a}) and the definition of momentum gives
the condition
$$2m_B u= \frac{1}{\gamma}2m_A u  \hspace*{6mm} \rightarrow \hspace*{4mm}m_A=\gamma m_B=\frac{1}{\sqrt{1-(v^A_x/c)^2}} m_B$$

\noindent
We may assume that $u$ is small so that $m_B$ is the \emph{mass at rest} of the ball, $m_0$, and $m_A$=$m(v^A)$ is the mass of the same 
ball when moving.

So we have found that 
 $$ m(v)=\gamma m_0$$
where  $m_0$ is the mass in the reference frame where the object is at rest.
We can keep the momentum definition, $\vec p$=$m \vec v$, from classical dynamics 
by giving up the invariance of mass. 
It is worth noting that in modern physics language $m$ is used for denoting the rest mass.
I will stick to the old notation for clarity.
An approach similar to Lewis and Tolman one is used in~\cite{RR68} where the elastic scattering of two identical particles
is observed in the center of mass and in the frame of one of the two particles.
The assumption done here (and in~\cite{RR68}) is that the scattering 
angle is equal to the incidence one which is a possible realization of an elastic scattering.

\newpage

\section{The relativistic energy}
As the mass depends upon $v$, let us write the Newton law $\vec F$=$m d\vec v/dt$ as

$$\vec F= \frac{d\vec p}{dt} =\frac{d }{dt}(\gamma m_0 \vec v)=m_0 \vec v \frac{d\gamma}{dt} + m_0\gamma \frac{d\vec v}{dt}
$$
\noindent
By scalar multiplication  of the r.hs. and l.h.s. by $ \vec{v}$, it is

$$ \vec F\cdot {\vec{v}}={\vec{v}}\cdot\frac{d\vec p}{dt}
$$
The l.h.s. is the work done by the force per unit time.
The r.h.s. gives
$$ {\vec{v}}\cdot\frac{d\vec p}{dt}= m_0\gamma{\vec{v}}\cdot\frac{d{\vec{v}}}{dt}+m_0\frac{v^3}{c^2}\gamma^3\frac{dv}{dt}=
m_0\gamma v\Bigl(1+\frac{v^2\gamma^2}{c^2}\Bigr)\frac{dv}{dt}=m_0\gamma^3v\frac{dv}{dt}
$$
that is
$$ \frac{dE}{dt}=\vec F\cdot {\vec{v}}=m_0\gamma^3v\frac{dv}{dt}$$
It is easy to verify that  this equation is satisfied if we define the energy as
$$E=mc^2=\gamma m_0c^2$$

For $v$=0, it is $E_0$=$m_0 c^2$ which has the meaning of the \emph{energy at rest}. Relativistically the
energy of a free particle at rest is non-vanishing.
\noindent
The (relativistic) kinetic energy is obtained by subtracting the rest energy from the total energy
$$T=mc^2-m_0c^2=m_0c^2(\gamma-1)\neq \frac{1}{2} \gamma m_0 v^2 $$ 
which gives the classical kinetic energy $T\simeq m_0v^2/2$ for $v\ll c$.

Experiments confirmed the validity of the relativistic relationship between $\vec p$ and $\vec v$.

\begin{figure}[htb]
\centering
\rotatebox{0}{
\includegraphics*[width=72mm]{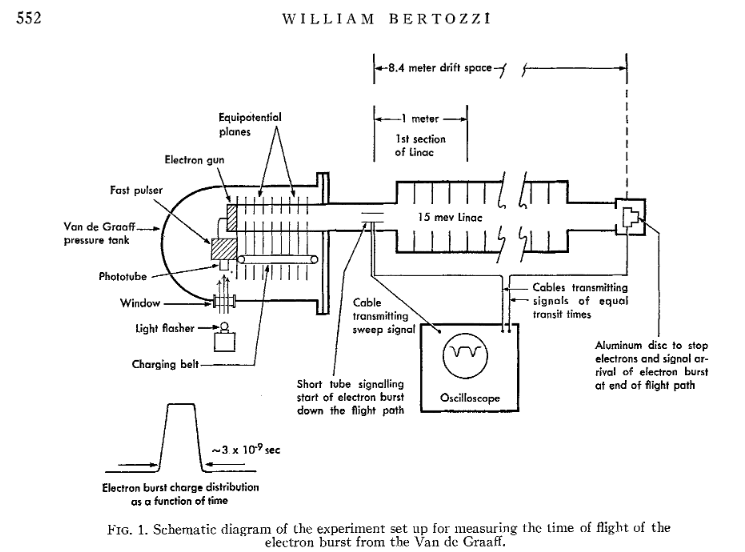}}
\caption{\label{eliana-rel-f15} Bertozzi apparatus (from~\cite{WB64}).}
\end{figure}

\begin{figure}[htb]
\centering
\rotatebox{0}{
\includegraphics*[width=72mm]{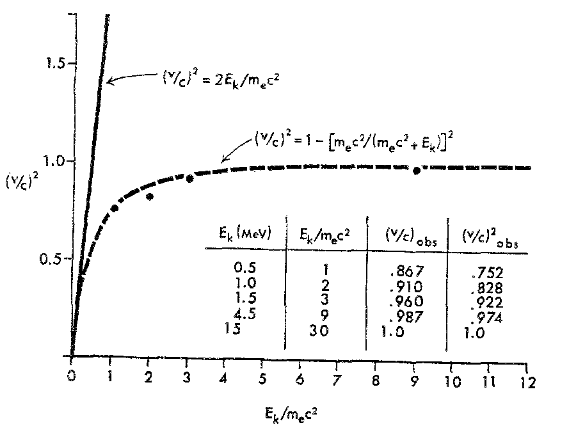}}
\caption{\label{eliana-rel-f16} Bertozzi results (from~\cite{WB64}). The solid line is the calculation using classical formulas, while the
dashed line is the relativistic prediction. The dots are the experimental results.}
\end{figure}

In particular Bertozzi experiment~\cite{WB64} measured directly the velocity of an $e^-$ beam  accelerated in a linear accelerator.
The experimental arrangement is shown in Fig.~\ref{eliana-rel-f15}.
The $e^-$ speed was measured through the time of flight. 
The kinetic energy was computed  from the accelerating field and from the measurement of
the heat deposited in the aluminum target.
\noindent
The results in Fig.~\ref{eliana-rel-f16} confirm Einstein prediction and also show clearly the presence of a limit speed, $c$. 

Relativity is of fundamental importance for accelerators where the particles may be accelerated to speed near to $c$.
In a ring accelerator dipole magnets keep the particles on the design orbit and longitudinal radio-frequency electric fields 
boost their energy.
The relationship between momentum and speed dictates how the frequency of the accelerating electric field and the 
dipole field must be varied with energy.

\begin{figure}[htb]
\centering
\rotatebox{0}{
\includegraphics*[width=68mm]{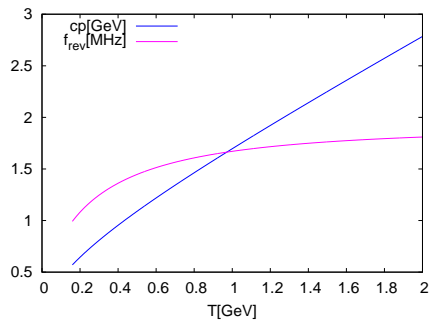}}
\caption{\label{eliana-rel-f17} Momentum and revolution frequency in the CERN PS Booster
as a function of the kinetic energy. The ring is about 157 m long.}
\end{figure}

The dipole field must be ramped up according to momentum for keeping the particles on the design orbit ($\rho$=$p/eB$).
The electric field frequency, which is a multiple of the revolution frequency, is 
$$f_{rf}= h f_{rev} = h\frac{\beta c} {L}=h\frac{c} { L}\sqrt{1-1/\gamma^2}   $$
which for large $\gamma$ becomes
$$ f_{rf} \approx h \frac{c}{L}\Bigl(1-\frac{1}{2\gamma^2}\Bigr) $$
At large $\gamma$ the revolution frequency is almost constant. This is
particularly true for $e^\pm$ which have 1836 larger $\gamma$  than protons for the same energy.
Fig.~\ref{eliana-rel-f17} shows the CERN PS Booster case where the protons kinetic energy is ramped from 160 MeV to 2 GeV.

\newpage
While in classical mechanics the mass is an invariant scalar conserved in physics processes, 
relativistically the rest mass alone is not conserved.

\begin{figure}[htb]
\begin{minipage}[c]{0.4\linewidth}
\rotatebox{0}{
\includegraphics*[width=56mm]{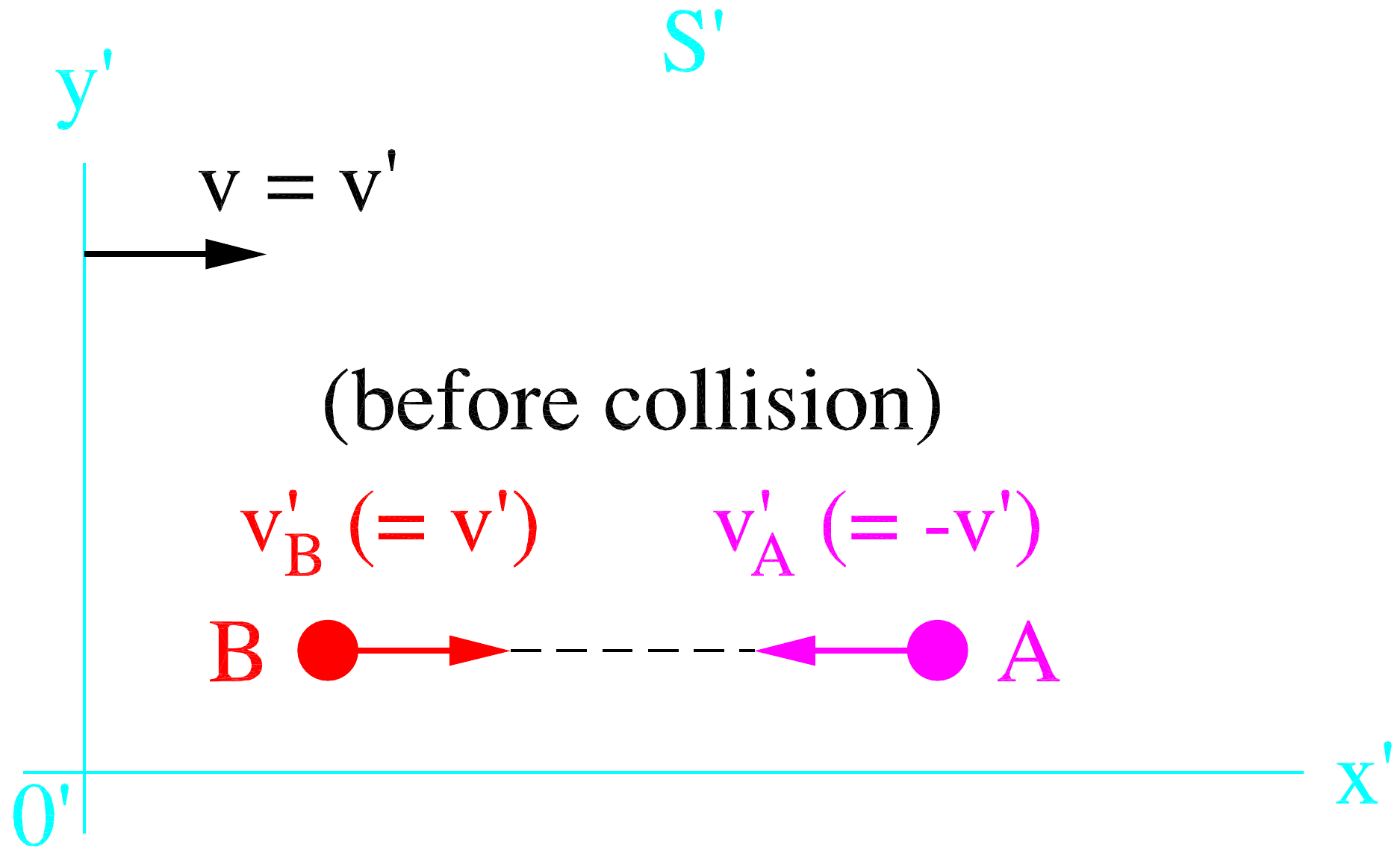}}
\caption{\label{eliana-rel-f18} Identical particle colliding head-on observed  in the center of mass frame , $S'$.}
\end{minipage}
\hfill
\begin{minipage}[c]{0.4\linewidth}
\rotatebox{0}{
\includegraphics*[width=56mm]{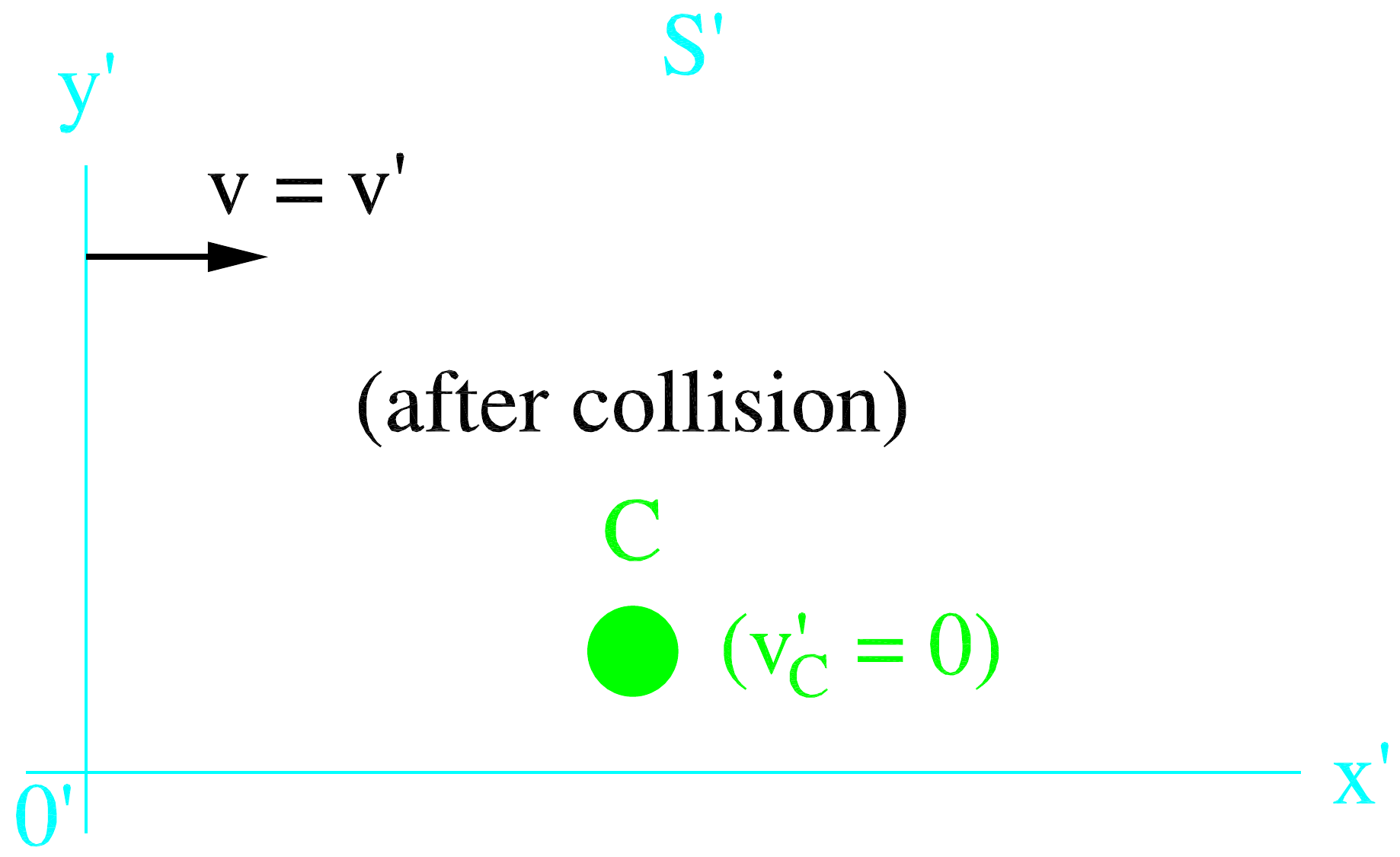}}
\caption{\label{eliana-rel-f19}  After inelastic collision the two particles glue together in the particle $C$ at rest in $S'$.}
\end{minipage}%
\end{figure}

\begin{figure}[htb]
\begin{minipage}[c]{0.4\linewidth}
\rotatebox{0}{
\includegraphics*[width=56mm]{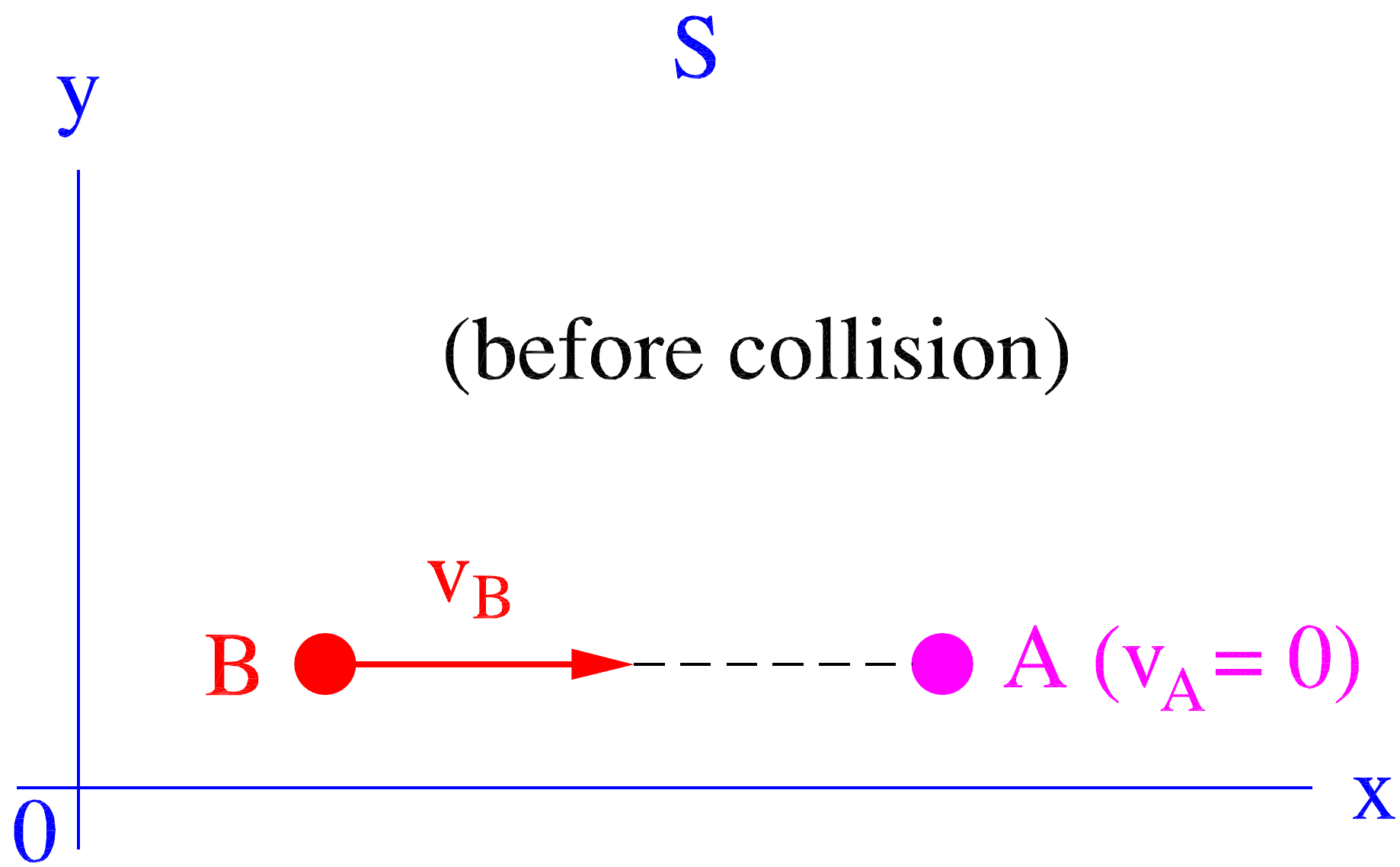}}
\caption{\label{eliana-rel-f20} Particles observed before collision in the frame $S$ where $A$ is at rest.}
\end{minipage}
\hfill
\begin{minipage}[c]{0.4\linewidth}
\centering
\rotatebox{0}{
\includegraphics*[width=56mm]{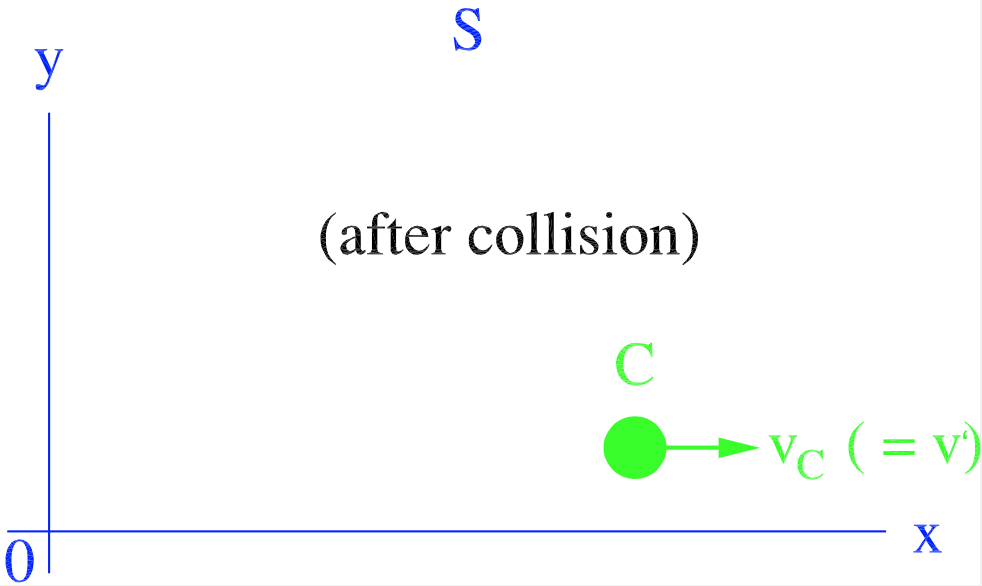}}
\caption{\label{eliana-rel-f21} Particle $C$ after collision as seen in $S$.}
\end{minipage}%
\end{figure}

To show that the rest mass is not conserved we consider an inelastic scattering (kinetic energy is not conserved) between two identical particles, $A$ and $B$, 
with rest mass $m_0$.
In the center of mass, $S'$, it is $\vec v'^A$=$-\vec v'^B$. We may assume $\vec v'^A$=$-\vec v'^B$=$\hat x v'$ 
(see Fig.~\ref{eliana-rel-f18}).
After colliding the two particles glue together in a new particle, $C$, at rest in $S'$ (see Fig.\ref{eliana-rel-f19}) so 
that momentum is conserved.  In the reference frame, $S$, where $A$ is at rest, the particle $B$ moves before the collision with speed
$v^B_x$=2$v'/(1+v'^2/c^2)$ while after the collision $C$ moves with speed $v^C_x$=$v'$ (see Figs.~\ref{eliana-rel-f20}, \ref{eliana-rel-f21}).
The mass of $B$ in $S$ is 
$$
m_B  = \frac{m_0}{\sqrt{1-(v^B_x/c)^2}}
      = \frac{m_0 [1+(v'/c)^2]}{1-(v'/c)^2} 
$$      
Momentum conservation in $S$ requires
\begin{gather*}
\begin{align*}
\underbrace{p^A_{x}+p^B_{x}}_{\text{before}} &  =
\underbrace{p^C_{x}}_{\text{after}}
\quad \rightarrow \quad
\frac{m_0 v^B_x}{\sqrt{1-(v_B/c)^2}} = \frac{m^C_0 v'}{\sqrt{1-(v'/c)^2}}
\end{align*}
\end{gather*}
Using the value found for $v^B_x$ and solving for $m^C_0$ we get
$$m^C_0 = \frac{2 m_0}{\sqrt {1-(v'/c)^2}}$$

$$m^C_0-2m_0 = 2m_0\biggl(\frac{1}{\sqrt{1-(v'/c)^2}}-1\biggr)$$
The rest mass of the product particle $C$  is \emph{larger} than the sum of the
starting particle rest masses and 
the difference, multiplied by $c^2$, is just the initial total kinetic
energy in $S'$, $T'_A+T'_B$.
The kinetic energy in $S'$ has been completely
converted into mass. Although the kinetic energy is not conserved the total
energy, kinetic plus energy at rest, is conserved.
The fact that the sum of the rest masses is not conserved is a
fact well known to every high energy particle physicist. An
example is the annihilation of a $e^+ e^-$ pair into 2 photons.

\section{Minkowski space-time and 4-vectors}
In 1907 the mathematician Hermann Minkowski, who was Einstein professor 
at Z\"urich Polytechnic, showed that the special theory of relativity 
can be formulated by using a 4-dimensional space with metric 
tensor\footnote{Here it is a 4$\times$4 matrix defining
the scalar product.}, $g$, given by

$$g= \left (
\begin{matrix} +1 &  0 & 0 & 0 \\
                        0  &-1 & 0 & 0 \\
                        0  & 0 & -1 & 0 \\                        
                        0  & 0 & 0 & -1 \\                        
                                  \end{matrix}
                                  \right )
$$
Lorentz frames are those frames where the metric tensor takes this special 
form; they are connected by Lorentz transformations.
The points of  Minkowski space-time are the \emph{events} and the vectors in this space have 4 components
which transforms according to Lorentz transformations (\emph{4-vectors}).

The transformations for momentum and energy may be found directly
from the definitions and the Lorentz transformations for the velocity.
The result is 
$$p'_x=\gamma_V(p_x-E \hspace*{1mm}V/c^2)   \hspace*{10mm} p'_y=p_y    \hspace*{10mm} p'_z=p_z$$
$$E'=\gamma_V (E-V\hspace*{1mm} p_x)$$
with $E$=$\gamma_vm_0c^2$ and $\gamma_V\equiv1/\sqrt{1-V^2/c^2}$.
\emph{A posteriori} we notice that the transformations have the same form as the Lorentz coordinates transformations with
$\vec r \rightarrow \vec p$ and $t \rightarrow E/c^2 $.

A more elegant way of reaching the same result is by noticing that 
$(E/c,\vec p)$ \emph{must} transform according to Lorentz transformations, and therefore it is a 4-vector.

While relativistically lengths and time depend upon the motion of the observer,  the interval defined as 

$$
\bigl(ds\bigr)^2 \equiv \bigl[d\bigl(ct\bigr)\bigr]^2
-\bigl(dx\bigr)^2-\bigl(dy\bigr)^2-\bigl(dz\bigr)^2
$$
is invariant under Lorentz transformations.
Indeed
\begin{align*}
ds'^2 & = c^2dt'^2-\bigl(dx'^2+dy'^2+dz'^2\bigr) \\
      & = \gamma^2\bigl(c^2dt^2+\beta^2dx^2
          -2\beta c \, dt \, dx-\beta^2 c^2dt^2-dx^2+2\beta c \, dt \, dx\bigr)-dy^2-dz^2 \\
      & = \gamma^2\bigl[\bigl(1-\beta^2\bigr)\bigl(c^2dt^2 -dx^2\bigr)\bigr]-dy^2-dz^2 \\
      & = c^2dt^2-dx^2 -dy^2-dz^2 = ds^2
\end{align*} 

\noindent
Let us consider a particle moving with velocity $\vec v(t)$, non necessarily uniform, in $S$.
The time interval $d\tau$ 
evaluated in a inertial frame $S'$
where the particle is
instantaneously at rest
is called proper time. 
It is
related to the time measured in $S$ by
\begin{align*}
d\tau & = \sqrt{1-v^2/c^2} dt \equiv \frac{dt}{\gamma} \\
\intertext{and for a finite time interval}
t_2-t_1 & = \int_{\tau_1}^{\tau_2} \frac{d\tau}{\sqrt{1-v^2/c^2}}
\end{align*}
The proper time is \emph{by definition} an invariant. This results also
from the fact that $c^2d\tau^2$ is
the invariant $ds^2$ evaluated in the frame
where the particle is instantaneously at rest.
This definition of proper time contains the definition given in Section 4 of Chapter 2
as a particular case when the particle is not accelerated.

\vspace*{4mm}
An object which 4 components transform as $(ct,x,y,z)$ is a 4-vector.
In the same way as done for  intervals, one can prove that for any 4-vector the quantity
\begin{align*}
A^\nu B_\nu & \equiv A_0 B_0 -\bigl(A_x B_x + A_y B_y +A_z B_z\bigr) 
\intertext{and in particular}
A^\nu A_\nu & = A_0^2 -\bigl(A_x^2 + A_y^2 + A_z^2\bigr) \nonumber
\end{align*}
are invariant. 

Classically the scalar products, $\vec A\cdot\vec B$, and in particular the length of vectors,
$\vec A\cdot\vec A$, are invariant. 

Owing to the fact that the proper time interval
$d \tau=dt/\gamma$ is an invariant and that
$(cdt,dx,dy,dz)$  transforms obviously as $(ct,x,y,z)$, the quantity (\emph{4-velocity}) defined as

\begin{eqnarray}\label{eliana-rel-eq12b}
\Bigl(\frac{c \, dt}{d\tau},\frac{dx}{d\tau},
\frac{dy}{d\tau},\frac{dz}{d\tau}\Bigr)  =\Bigl(\gamma\frac{c \, dt}{dt},\gamma\frac{dx}{dt},
\gamma\frac{dy}{dt},\gamma\frac{dz}{dt}\Bigr)  \equiv
(\gamma c,\gamma \vec{v})
\end{eqnarray}

transforms according to Lorentz transformations. Multiplying the 4-velocity by  the rest mass we get

$$
m_0(\gamma c,\gamma \vec{v})=(E/c,\vec p)
$$
which is also a 4-vector (energy-momentum 
or 4-momentum vector). Therefore it
transforms according to Lorentz transformation and
the quantity $(E/c)^2 -(p_x^2+p_y^2+p_z^2)$ is an invariant. 

Relativistically energy and momentum are closely connected.
If in one inertial reference frame energy and momentum are conserved ($\Delta \vec p$=0 and $\Delta E$=0),
for example in a collision between
particles,  they are conserved
in every other inertial frame because a
4-vector having all components vanishing in a reference frame
will have vanishing components in any other one too.

Similarly if momentum is
conserved for two inertial observers ($\Delta \vec p$=$\Delta {\vec p}\hspace*{1mm}'$=0), the energy too must be conserved.

\section{ Newton and Minkowski force and their relativistic transformation}

We may  write the relativistic Newton law $\vec F$=$d\vec p/dt$  in terms of 4-vectors. In the particle proper frame

\begin{eqnarray}\label{eliana-rel-eq13}
\frac{dp^\nu}{d\tau}=f^\nu
\end{eqnarray}
with $(p^0,p^1,p^2,p^3)$=$(E/c, p_x,p_y,p_z)$ and $(f^0,f^1,f^2,f^3)$=$(f^0,F_x,F_y,F_z)$.
The l.h.s. is a 4-vector and therefore also $\vec f$, the Minkowski force,  on the r.h.s. must be a 4-vector
related to the Newton force $\vec F$. 

The space part of the equation of motion is
 
\begin{gather*}
 \frac{d \vec{p}}{d \tau} = \vec{f}
\qquad \rightarrow  \qquad
\gamma \frac{d \vec{p}}{dt} = \vec{f} 
\qquad \rightarrow  \qquad
\vec{f} =\gamma \vec{F}%
\end{gather*}

The time part of Eq.~(\ref{eliana-rel-eq13}) is

\begin{eqnarray}\label{eliana-rel-eq14}
f^0 = \frac{d p^0}{d \tau}
    = \frac{1}{2p^0} \frac{d (p^0)^2}{d \tau}=\frac{1}{2p^0} \frac{d (E/c)^2}{d \tau}
\end{eqnarray}

The invariance of $(E/c)^2 -\vec p\cdot\vec p=(m_0c)^2$
implies that

$$
\frac{d}{d \tau} \biggl[ \left(\frac{E}{c}\right)^2 -\vec{p} \cdot \vec{p} \biggr]=0
$$
$$
 \frac{d}{d \tau} \left(\frac{E}{c}\right)^2
= 2\vec{p} \cdot \frac{d \vec{p}}{d \tau}
$$
which inserted in Eq.~(\ref{eliana-rel-eq14}) gives

\begin{eqnarray}
f^0 = 
\frac{1}{2p^0} \frac{d (E/c)^2}{d \tau}
   = \frac{1}{2 p^0} 2 \vec{p} \cdot \frac{d \vec{p}}{d \tau}
    = \frac{m_0 \gamma\vec{v}}{E/c}\cdot \bigl(\gamma \vec{F}\bigr)
   = \gamma \vec{\beta} \cdot  \vec{F} 
\end{eqnarray} 

The Minkowski force is therefore
$$(f^0,f^1,f^2,f^3)=(\gamma\vec\beta\cdot \vec F, \gamma \vec F)$$

We notice that

\begin{alignat*}{3}
 \frac{dE}{dt} = \frac{1}{\gamma}\vec{v} \cdot \vec{f}
              &  = \frac{1}{\gamma}\frac{d\vec{\ell}}{dt} \cdot \vec{f}
                 \qquad &  \rightarrow &&
\hspace*{4mm}dE &  = \frac{1}{\gamma} d \vec{\ell} \cdot \vec{f}
\end{alignat*}

which is the expression of the work done by a force $\vec{F}=\vec{f} /\gamma$.

In absence of external forces ($\vec{F}$=0) it is $\vec f$=0  and momentum and energy
are conserved.



Being a 4-vector, Minkowski force transforms following Lorentz transformations. It must be paid attention to distinguish between the
particle velocity, $\vec v$, in the $S$ frame and the frames relative speed that we will denote by $\vec V$. 
Using the general expression of Lorentz transformations Eq.~(\ref{eliana-rel-eq5a}) we have

\begin{align*}
f'^0 & = \gamma_V \bigl(f^0 - \vec{\beta}_V \cdot \vec{f}\bigr) \\
\vec{f}\hspace*{1mm}' & = \vec{f} +\frac{\gamma_V-1}{\beta_V^2}
 \bigl(\vec{\beta}_V \cdot \vec{f}\,\bigr) \vec{\beta}_V-\gamma_V f^0
\vec{\beta}_V
\end{align*}
The Newton force transformation writes

\begin{equation}\label{eliana-rel-eq16}
\gamma\hspace*{0.5mm}' \vec{F}' = \gamma \vec{F} +\frac{\gamma_V-1}{\beta_V^2}
 \bigl[\vec{\beta}_V \cdot \bigl(\gamma\vec{F}\bigr)\bigr] \vec{\beta}_V-\gamma_V
 \vec{\beta}_V \bigl(\gamma \vec{\beta} \cdot  \vec{F}\bigr) 
 \end{equation} 
The inverse transformation is obtained by replacing
$\vec{\beta}_V$ with $-\vec{\beta}_V$.

For $V\ll c$ ($\beta_V\rightarrow$ 0 and $\gamma_V\rightarrow$ 1) it is $\vec F'$=$\vec F$ which is the classical result.

For the translational motion along $x$ the transformations write

\begin{equation}\label{eliana-rel-eq16a}
F'_x=F_x-\frac{v_y V}{c^2-v_xV} F_y-\frac{v_z V}{c^2-v_xV} F_z  \hspace*{14mm}
F'_{y,z}=\frac{\sqrt{1-V^2/c^2}}{1-v_xV/c^2} F_{y,z}
\end{equation}

If the force $\vec F$ is acting on a particle which is instantaneously at rest in $S$ ($v$=0), the transformations simplify
\begin{equation}\label{eliana-rel-eq16b}
F'_x=F_x \hspace*{6mm}  F'_y= \frac{1}{\gamma} F_y \hspace*{6mm}  F'_z= \frac{1}{\gamma} F_z
\end{equation}

\section{Some geometrical aspects of special relativity}

If $(ct_1,x_1,y_1,z_1)$ and $(ct_2,x_2,y_2,z_2)$ are the coordinates of two events in $S$
we ask whether it is possible to find an inertial frame $S'$ where the two events happen in the \emph{same place}.
As the interval $ds^2$  is invariant (see Section 3 of Chapter 3) this means that

$$(\Delta s')^2 =  (\Delta s)^2$$
and therefore
$$(c\Delta t')^2 = (c\Delta t )^2-(\Delta x^2 +\Delta y^2 +\Delta z^2)$$
where we have set  $\Delta t \equiv t_2-t_1$, $\Delta x \equiv x_2-x_1$ and so on.
The l.h.s. of this equation is always positive.
Therefore the answer is affirmative if $(\Delta s)^2 >$0. Such intervals are called \emph{time-like} intervals.
The time in $S'$ between the two events is 
$$\Delta t'=\frac{1}{c}\sqrt{c^2\Delta t^2-(\Delta x^2 +\Delta y^2 +\Delta z^2)}=\frac{\Delta s}{c}$$
For the simple case of Eqs.(\ref{eliana-rel-eq5})
we find that the speed of the frame $S'$ with respect to $S$ 
is $V$=$\Delta x/\Delta t$ which is smaller than $c$ because we have assumed 
$(\Delta s)^2 >$0. 

Now we ask if it is possible to find an inertial frame where the two events happen at the \emph{same time}.

In this case $(\Delta s')^2 =  (\Delta s)^2$ implies that 
$$  (c\Delta t )^2-(\Delta x^2 +\Delta y^2 +\Delta z^2)= -(\Delta x'^2 +\Delta y'^2 +\Delta z'^2) < 0$$
that is $(\Delta s)^2$ must be negative.
The distance between the two events in $S'$ is
$$\sqrt{\Delta x'^2 +\Delta y'^2 +\Delta z'^2}=\sqrt{\Delta x^2 +\Delta y^2 +\Delta z^2-(c\Delta t)^2 }
$$
which is a \emph{real} number as the argument of the square root on the l.h.s is positive.

By using the Lorentz transformations Eqs.(\ref{eliana-rel-eq5}) we find 
$$0=c\Delta t'=\gamma(c\Delta t-\beta\Delta x)$$
that is the speed $V$ of the frame $S'$ with respect to $S$  is  $V$=$c^2\Delta t/\Delta x$. The constraint
$v<c$ imposes $\Delta x/\Delta t>c$. This means that between the two events there 
may exist no causality connection.
These intervals are called \emph{space-like} intervals.
\vspace*{0mm}
\noindent
Finally the case $\Delta s'$=$\Delta s$=0 corresponds to events connected by a light ray.

\begin{figure}[htb]
\centering
\rotatebox{0}{
\includegraphics*[width=56mm]{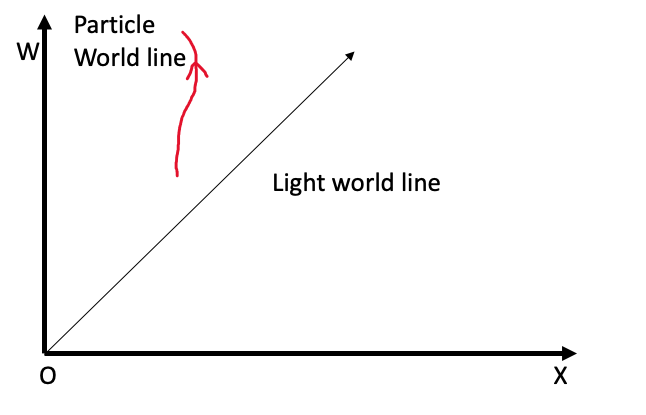}}
\caption{\label{eliana-rel-f22} $(x,w)$ diagram relative to the inertial reference frame $S$.}
\end{figure}

Let us consider our observer $O$ at the origin of the inertial frame $S$. 
We can represent the $x$ and $w\equiv ct$ coordinates\footnote{For simplicity only the space coordinate $x$
is considered.}
measured by $O$  on two orthogonal axis (see Fig.~\ref{eliana-rel-f22}). This graphical illustration was introduced
by  Minkowski. 
Any event is represented by a point in the Minkowski diagram
and  the trajectory of a particle will be a sequence of points called ``world line".
The angle between the tangent to a material particle world line and the $w$-axis is always smaller than 45$^0$,
as the particle speed is always smaller than $c$. The world line of a light ray is a 
straight line at 45$^0$.

\begin{figure}[htb]
\centering
\rotatebox{0}{
\includegraphics*[width=56mm]{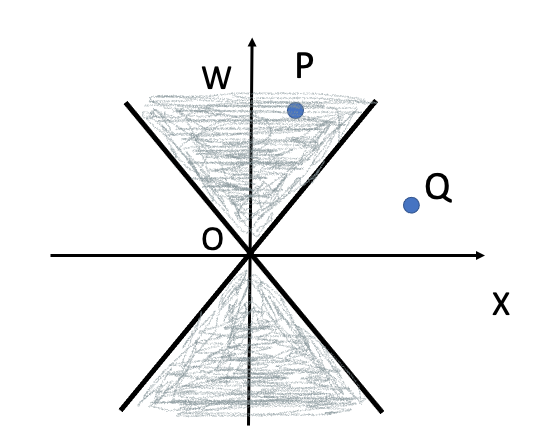}}
\caption{\label{eliana-rel-f24} The light cone relative to the observer $O$. $P$ is an event in the future, while $Q$ is an 
``elsewhere'' event.}
\end{figure}

Let us consider the $(x,w)$ diagram relative to an inertial reference frame $S$. The world lines 
of light waves delimit the grey area in Fig.~\ref{eliana-rel-f24} and define the so called 
\emph{light cone}. For any  event point inside the grey area, $P$,  it is $w^2-x^2>$0. That is the interval $\Delta s$
between those points and $O$ are time-like and it is always
possible to find a Lorentz transformation where the event happens in the same place and therefore it can be
established their chronological sequence. The events in the upper part of the grey region for which $t>$0
happen after the event $O$. This region is called \emph{future} with respect to $O$ ). The events 
represented by points in the lower part of the grey region for which  $t<$0 happen before $O$ (\emph{past}). 
As the interval $\Delta s^2$ is invariant the fact that  $P$ is a future event with respect to $O$
does not depend upon the reference frame.

All points like $Q$ outside the grey area correspond to space-like intervals because $\Delta s^2$=$(ct )^2-x^2<$0.
As previously shown these are space-like intervals for which it is not possible to find a reference frame
where the events happen in the same space point. Therefore it is not possible to establish a chronological sequence between 
them. This region is called \emph{elsewhere}.

\chapter{Relativistic transformations of EM fields and sources}
\section{Relativistic transformations of EM fields}
The force acting on a charged particle moving in a EM field with velocity $\vec v$ is the Lorentz force
$$
\vec F = q \vec E + q \vec v \times \vec B 
$$
The corresponding  Minkowski force is
\begin{align*}
f^\nu & = \bigl(\gamma \vec{\beta} \cdot \vec{F}, \gamma \vec{F}\bigr)
        = q \, \bigl[\gamma \vec{\beta} \cdot \bigl(\vec{E} + \vec{v} \times \vec{B}\bigr),
          \gamma \, \bigl(\vec{E} + \vec{v} \times \vec{B}\bigr)\bigr] =q \, \bigl[\gamma \vec{\beta} \cdot \vec{E},
          \gamma \, \bigl(\vec{E} + \vec{v} \times \vec{B}\bigr)\bigr] 
\end{align*}    
with $\gamma=1/\sqrt{1-(v/c)^2}$.    
This equation can be written in matrix form  as
$$
\left (
\begin{matrix} f^0 \\
                       f^1\\
                       f^2\\                        
                       f^3 \\                        
                                  \end{matrix}
                                  \right )=\frac{q}{c}
                                  \left (
\begin{matrix}  0 &  E_x & E_y & E_z \\
                        E_x  & 0 & cB_z & -cB_y \\
                        E_y & -cB_z & 0 & cB_x \\                        
                        E_z  & cB_y & -cB_x & 0 \\                        
                                  \end{matrix}
                                  \right )
                                  \left (
\begin{matrix} \gamma c \\
                       \gamma v_x\\
                       \gamma v_y \\                        
                       \gamma v_z\\                        
                                  \end{matrix}
                                  \right )
                                  $$
In the moving frame $S'$ the Minkowski force will be expressed in the same form
in terms of the primed quantities.                                     
The Minkowski force and the 4-velocity $(\gamma c, \gamma \vec v)$ (Eq.~\ref{eliana-rel-eq12b}) are 4-vectors.                                 
Using the Lorentz transformation  ${\cal L}$  from $S$ to $S'$ and ${\cal L}^{-1}$ from $S'$ to $S$ we get 
$$
\left (
\begin{matrix} f'^0 \\
                       f'^1\\
                       f'^2\\                        
                       f'^3 \\                        
                                  \end{matrix}
                                  \right )=
{\cal L}
\left (
\begin{matrix} f^0 \\
                       f^1\\
                       f^2\\                        
                       f^3 \\                        
                                  \end{matrix}
                                  \right )=\frac{q}{c}{\cal L}
                                  \left (
\begin{matrix}  0 &  E_x & E_y & E_z \\
                        E_x  & 0 & cB_z & -cB_y \\
                        E_y & -cB_z & 0 & cB_x \\                        
                        E_z  & cB_y & -cB_x & 0 \\                        
                                  \end{matrix}
                                  \right )
                            {\cal L}^{-1}      \left (
\begin{matrix} \gamma' c \\
                       \gamma' v'_x\\
                       \gamma' v'_y \\                        
                       \gamma' v'_z\\                        
                                  \end{matrix}
                                  \right )
                                  $$

Requiring that the Minkowski force in $S'$ has the same form as  in $S$,  it must be
$$
 \left (
\begin{matrix}  0 &  E'_x & E'_y & E'_z \\
                        E'_x  & 0 & cB'_z & -cB'_y \\
                        E'_y & -cB'_z & 0 & cB'_x \\                        
                        E'_z  & cB'_y & -cB'_x & 0 \\                        
                                  \end{matrix}
                                  \right )=
{\cal L}
                                  \left (
\begin{matrix}  0 &  E_x & E_y & E_z \\
                        E_x  & 0 & cB_z & -cB_y \\
                        E_y & -cB_z & 0 & cB_x \\                        
                        E_z  & cB_y & -cB_x & 0 \\                        
                                  \end{matrix}
                                  \right )
                            {\cal L}^{-1}     
$$

which yelds  the field components in $S'$~\cite{HH}
\begin{alignat*}{2}
E'_x & = E_x & \qquad
B'_x & = B_x \\
E'_y & =\gamma_V (E_y-VB_z) & \qquad
B'_y & = \gamma_V \Bigl(B_y +{V\over c^2} E_z\Bigr) \\
E'_z & =\gamma_V (E_z+VB_y) & \qquad
B'_z & = \gamma_V \Bigl(B_z -{V\over c^2} E_y\Bigr) \\
\end{alignat*}


In alternative to the previous formal approach,
we give here a way for finding directly the
field transformation from physical considerations~\cite{RR68}.

The Minkowski force associated to the Lorentz force in $S$  is
\begin{align*}
f^\nu & = \bigl(\gamma \vec{\beta} \cdot \vec{F}, \gamma \vec{F}\bigr)
        = q \, \bigl[\gamma \vec{\beta} \cdot \vec{E} ,
          \gamma \, \bigl(\vec{E} + \vec{v} \times \vec{B}\bigr)\bigr] 
\end{align*}       
In a second reference frame, $S'$,
           the force must have the same form
\begin{align*}           
f'^\nu & = \bigl(\gamma' \vec{\beta}' \cdot \vec{F}', \gamma' \vec{F}'\bigr)
= q \, \bigl[\gamma' \vec{\beta'} \cdot \vec{E'} ,
          \gamma' \, \bigl(\vec{E'} + \vec{v'} \times \vec{B'}\bigr)\bigr] 
\end{align*}
where we assumed $q'=q$ which is a fact experimentally proven with high
precision.

Knowing how the Minkowski force transforms it is possible to get the
expressions for the field transformations.

Let us consider the
case of a particle at rest in $S$ subject to the
fields $\vec E$ and $\vec B$.
In $S$ it is
$$
\vec F = q \vec E
$$
In the frame $S'$ moving with translational motion along
the common $x$-axis with velocity $V$ with respect to $S$ it is 
$v_x'$=$-V$ and $v'_y$=$v'_z$=0.
The force components in $S'$ are
\begin{align*}
F'_x & = q(E'_x+v'_yB'_z-v'_zB'_y)=qE'_x \\
F'_y & = q(E'_y-v'_xB'_z+v'_zB'_x)=qE'_y+qVB'_z \\
F'_z & = q(E'_z+v'_xB'_y-v'_yB'_x)=qE'_z-qVB'_y
\end{align*}
From Eq.~(\ref{eliana-rel-eq16b}) 
the force components transform as \begin{align*}
F'_x & = F_x  \\
\gamma_V F'_y & = F_y \\
\gamma_V F'_z & = F_z
\end{align*}
with $\gamma_V\equiv1/\sqrt{1-(V/c)^2}$.
Writing explicitly the force in terms of the fields we get

\begin{alignat*}{3}
E_x & = E'_x & \qquad
E_y & =\gamma_V (E'_y+VB'_z) & \qquad
E_z & =\gamma_V (E'_z-VB'_y) \\
\end{alignat*}

The
inverse transformation are obtained replacing $V$ with $-V$
\begin{alignat*}{3}
E'_x & = E_x  & \qquad
E'_y = \gamma_V (E_y-&VB_z) & \qquad
E'_z  = \gamma_V (E_z+&VB_y)
\end{alignat*}

Finding out
the transformation for the magnetic field is a more complicated
because the electric force cannot be made vanishing by a convenient
choice of the reference frame.
We consider again the two frames $S$ and $S'$, with
$S'$ moving with velocity $V$ along the common $x$-axis.
For a charged particle moving in $S'$ along the $y'$-axis
it is $v_x$=$V$, $v_y$=$v'_y/\gamma_V$ and $v_z$=$v'_z$=0.
The force in $S'$ is 
$$
F'_x = q(E'_x+v'_yB'_z) \qquad
F'_y = qE'_y \qquad
F'_z = q(E'_z-v'_yB'_x)
$$
Using the force transformation we get
\begin{align*}
\gamma'F'_x & = \gamma' q(E'_x+v'_yB'_z)
              = \gamma F_x+ {\gamma_V -1
                \over \beta_V^2 } \beta_V^2\gamma F_x -\gamma_V \gamma
                \beta_V({v_x \over c}F_x+{v_y \over c} F_y) \\
            & = {\gamma \over \gamma_V}F_x-\gamma_V \gamma{v_y V \over c^2} F_y
              = {\gamma \over \gamma_V}q(E_x+v_yB_z)-\gamma_V \gamma{v_y V
                \over c^2} q(E_y-v_xB_z) \\
\gamma'F'_y & = \gamma' qE'_y
              = \gamma F_y
              = \gamma q(E_y-v_xB_z) \\
\gamma'F'_z & = \gamma' q(E'_z-v'_yB'_x)
              = \gamma F_z
              = \gamma q(E_z+v_xB_y-v_yB_x)
\end{align*}

Using the transformations already found for the electric field and the fact that
in this case it is $\gamma' \gamma_V=\gamma$,
we notice that the equation for $F'_y$ is an identity while 
the other two equations give
$$
\gamma v_y B'_z = {\gamma \over \gamma_V} v_yB_z-
                  \gamma \gamma_V {v_y V \over c^2}(E_y-VB_z)
$$
$$
\gamma' \gamma_V(E_z+VB_y)-\gamma_Vv_yB'_x
                 = \gamma (E_z+VB_y-v_yB_x)
$$
The magnetic field component transformations are therefore
\begin{align*}
B'_z & = \gamma_V (B_z -{V\over c^2} E_y) \\
B'_x & = B_x 
\end{align*}

The transformation for $B_y$ is obtained considering
a particle moving along the $z'$-axis and writes
\begin{align*}
B'_y & = \gamma_V (B_y +{V\over c^2} E_z)
\end{align*}


The expressions found are valid for a translational motion along the $x$-axis.
In the general case when $\vec V$ has an arbitrary direction the field transformations write~\cite{DJ75}
\begin{equation}
\begin{aligned}
\vec{E}' & = \gamma_V\bigl(\vec{E}+\vec{V} \times \vec{B}\bigr) -\frac{\gamma_V^2}{\gamma_V+1}
             (\vec{\beta}_V \cdot \vec{E}) \vec{\beta}_V \\
\vec{B}' & = \gamma_V\bigl(\vec{B} -\frac{\vec{V}}{c^2} \times \vec{E}\bigr) -
             \frac{\gamma_V^2}{\gamma_V+1}\bigl(\vec{\beta}_V \cdot \vec{B}\bigr) \vec{\beta}_V
\end{aligned}
\end{equation}


Decomposing the fields in their components parallel and perpendicular
to the relative velocity $\vec V$, these relations may be written also as
\begin{align*}
\vec{E}' & = \vec{E}_{\parallel}+\gamma_V\bigl(\vec{E}_\bot
           + \vec{V} \times \vec{B}\bigr) \\
\vec{B}' & = \vec{B}_{\parallel}+\gamma_V\bigl(\vec{B}_\bot
           - \frac{\vec{V}}{c^2}\times \vec{E}\bigr)
\end{align*}
where we made use of the identity
$$
\gamma_V-\frac{\gamma^2_V \beta^2_V}{\gamma_V +1} = 1
$$

\section{Transformation of a charge distribution}
\begin{figure}[htb]
\centering
\rotatebox{0}{
\includegraphics*[width=42mm]{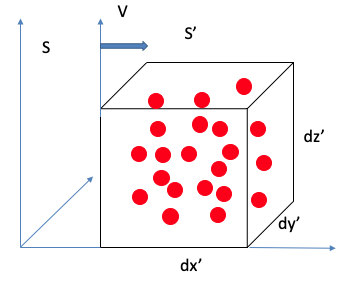}}
\caption{\label{eliana-rel-f25} Charge distribution at rest in $S'$ .}
\end{figure}

Let us consider a distribution of charges at rest in $S'$.
The charge density is given by
$$
\rho'(x',y',z',t') = \frac{qN}{dx' dy' dz'}
$$
In the $S$ frame which moves with velocity $-V$ with respect to $S'$ (see Fig.~\ref{eliana-rel-f25}),
the volume element is

$$dx\,  dy \, dz= \frac{dx'}{\gamma} dy' dz'
$$
where we have taken into account the length contraction in the $x$ direction.

The charge density in $S$ is therefore
\begin{gather*}
\rho = \frac{qN}{dx\, dy\, dz}
     = \gamma \rho' = \gamma \rho_0
 \end{gather*}
where we have renamed with $\rho_0$ the charge density in the rest frame, $\rho'$.
As the charge distribution moves in $S$ with velocity $+\hat x V$, in $S$  there is also a current moving in the $x$ direction with density

\begin{gather*}
j_x = \rho V
    = \gamma \rho' V 
\end{gather*}
and in general 
$$\vec{j} = \rho\vec{V}
                                  = \gamma\rho'\vec{V}$$

Multiplying the 4-velocity by the charge density at rest $\rho_0$ we get the 4-vector
$$\rho_0 (\gamma c, \gamma \vec V)=(\rho c,\rho \vec V)=(\rho c, \vec j)$$ 
(charge-current 4-vector).  Indeed the transformations we have found are the (inverse) Lorentz transformations for the particular case $\vec j'$=0.                              


\section{Potential 4-vector}
\noindent
In the Lorentz gauge

$$\nabla \cdot\vec A=- \frac{1}{c^2}\frac{\partial \Phi}{\partial t} $$ 
the equations for the scalar and vector potential take the form 
$$
\frac{1}{c^2}\frac{\partial^2 \Phi}{\partial t^2} - \nabla^2\Phi = \frac{\rho}{\epsilon_0}
$$
$$
\frac{1}{c^2}\frac{\partial^2 \vec A}{\partial t^2} - \nabla^2\vec A = \frac{\vec j}{\epsilon_0c^2}
$$
Using the d'Alembert operator
$$\Box \equiv \frac{1}{c^2}\frac{\partial^2}{\partial t^2} - \nabla^2 $$
these equations can be combined in a single one

\begin{equation}{\label{eliana-rel-eq19}}
\Box A^\alpha = \frac{1}{\epsilon_0c^2}\mu_0 J^\alpha
\end{equation}
with  $ A^0$=$\Phi/c$, $A^1$=$A_x$, $A^2$=$A_y$, $A^3$=$A_z$ and  $J^0$=$c\rho$, $J^1$=$j_x$,
$J^2$=$j_y$, $J^3$=$j_z$. We know now that $(c\rho,\vec j \hspace*{1mm})$ is a 4-vector and it is easy to verify that 
the d'Alembert operator is invariant under Lorentz transformations. Therefore
also $(\Phi/c,\vec A)$ must be a 4-vector.

\section{Direct proof of invariance of Maxwell equations}

Knowing how fields and sources
transform one can prove that Maxwell equations
are invariant under Lorentz transformation.

\noindent
For example let us prove that
$$
\nabla \cdot \vec{E} = \frac{\rho}{\epsilon_0}
\qquad \Rightarrow \qquad
\nabla' \cdot \vec{E}' = \frac{\rho'}{\epsilon_0}
$$
The partial
derivatives in $S'$ and in $S$ are related by the cyclic rule
\begin{alignat*}{2}
\frac{\partial}{\partial ct'} & = \frac{\partial ct}{\partial ct'}
\frac{\partial}{\partial ct}+\frac{\partial x}{\partial ct'}
\frac{\partial}{\partial x}+\frac{\partial y}{\partial ct'}
\frac{\partial}{\partial y}+\frac{\partial z}{\partial ct'}
\frac{\partial}{\partial z}
  & = \gamma \left(\frac{\partial}{\partial ct}+
  \beta\frac{\partial}{\partial x} \right) & \\
\frac{\partial}{\partial x'} & =
\frac{\partial ct}{\partial x'}\frac{\partial}{\partial ct}+
\frac{\partial x}{\partial x'}\frac{\partial}{\partial x}+
\frac{\partial y}{\partial x'}\frac{\partial}{\partial y}+
\frac{\partial z}{\partial x'}\frac{\partial}{\partial z}
  & = \gamma\left(\beta\frac{\partial}{\partial ct} +
  \frac{\partial}{\partial x}\right) &
\end{alignat*}
$$
\frac{\partial}{\partial y'} = \frac{\partial}{\partial y}
\qquad\qquad
\frac{\partial}{\partial z'} = \frac{\partial}{\partial z}
$$
By using the cyclic rule, the EM field transformations 
and the fact that Maxwell equation hold good in $S$, we find 
\begin{align*}
\nabla' \cdot \vec{E}' & = \frac{\partial E'_x}{\partial x'}
                          +\frac{\partial E'_y}{\partial y'}
                          +\frac{\partial E'_z}{\partial z'} \\
                       & = \gamma\frac{\partial E'_x}{\partial x}
                          +\frac{\partial E'_y}{\partial y}
                          +\frac{\partial E'_z}{\partial z}
                          +\gamma\beta \frac{\partial E'_x}{\partial ct} \\
                       & = \gamma \frac{\partial E_x}{\partial x}
                          +\gamma\frac{\partial E_y}{\partial y}
                          +\gamma\frac{\partial E_z}{\partial z}
                          -\gamma V\frac{\partial B_z}{\partial y}
                          +\gamma V\frac{\partial B_y}{\partial z}
                          +\gamma\beta\frac{\partial E_x}{\partial ct} \\
                       & = \gamma \nabla \cdot \vec{E}
                          -\gamma V \left(\frac{\partial B_z}{\partial y}
                          -\frac{\partial B_y}{\partial z}\right)
                          +\gamma\beta\frac{\partial E_x}{\partial ct} \\
                       & = \gamma\frac{\rho}{\epsilon_0}
                          -\gamma V \biggl(\nabla \times \vec{B}
                          -\frac{1}{c^2}\frac{\partial \vec{E}}{\partial t}\biggr)_x \\
                       & = \gamma\frac{\rho}{\epsilon_0}
                          -\gamma V\frac{j_x}{\epsilon_0 c^2} \\
                       & = \frac{\gamma}{\epsilon_0 c} \left(\rho c
                          - \beta j_x \right) \\
                       &= \frac{\rho'}{\epsilon_0}
\end{align*}
which prove the invariance of the first Maxwell law under Lorentz transformations.

\chapter{Some simple applications of EM transformations}
The fact that physics laws are the same in any reference frame allows us to solve problems in the most convenient reference frame.
Here we show two typical examples which are relevant in accelerator physics. 

\section{The field of a moving charge}
The EM fields generated by a moving charge have a simple form in the frame $S'$ 
where the particle is at rest, namely

\begin{align*}
\vec{E}' & = \frac{q}{4 \pi \epsilon_0} \frac{\vec{r}'}{r'^3} \\
\vec{B}' & = 0 
\end{align*}

\begin{figure}[htb]
\centering
\rotatebox{0}{
\includegraphics*[width=42mm]{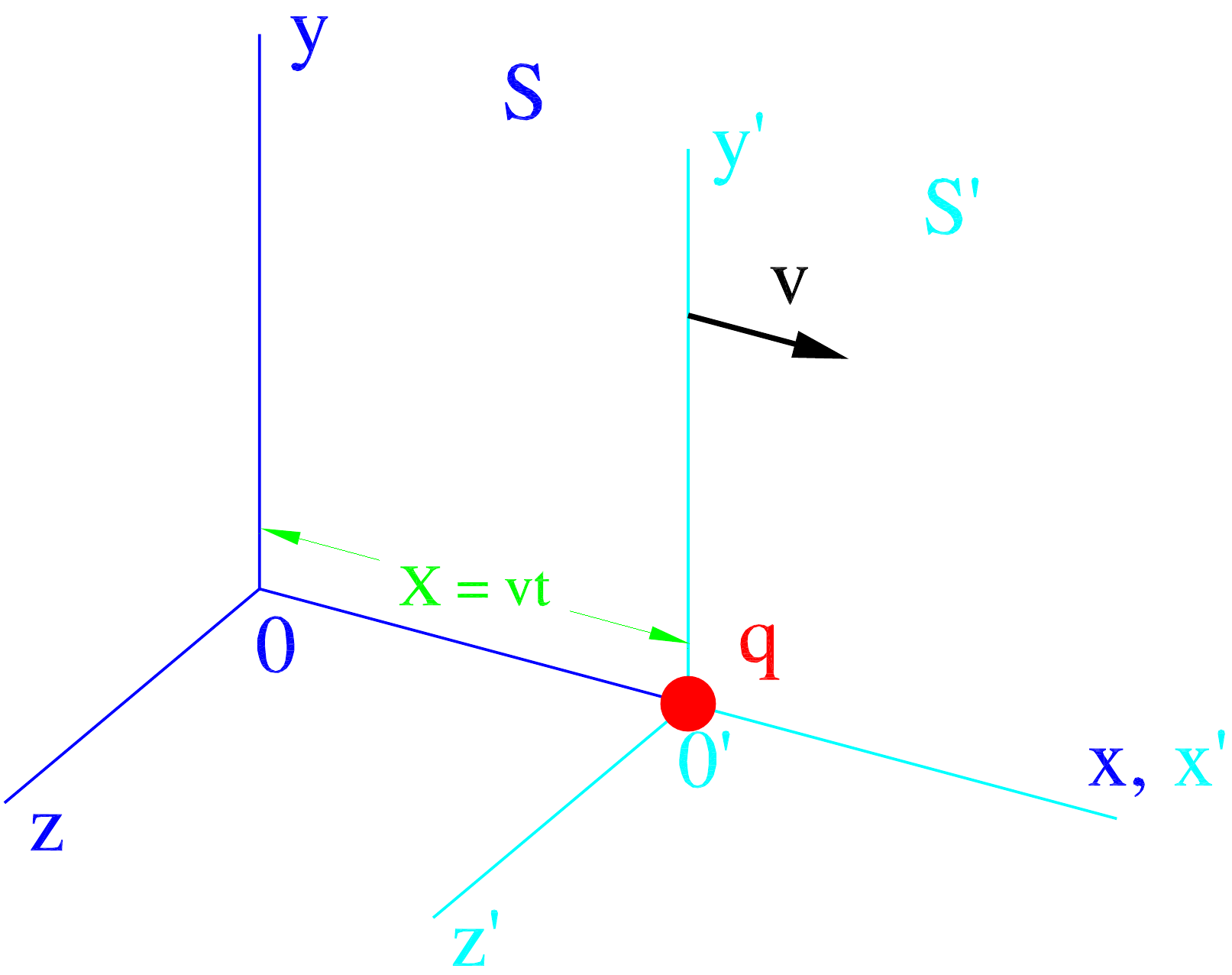}}
\caption{\label{eliana-rel-f26} Particle moving along the $x$-axis of reference $S$.}
\end{figure}

We can use the field transformations found in Chapter 4 for computing the fields in the frame where 
the particle is uniformly moving. We chose the frame so that the particle moves along the $x$-axis (see Fig.~\ref{eliana-rel-f26}).

The electric field in $S$ is
\begin{alignat*}{3}
E_x = E'_x
    & = \frac{q}{4 \pi \epsilon_0} \frac{x'}{r'^3}  & \qquad
E_y = \gamma E'_y
    &  = \gamma\frac{q}{4 \pi \epsilon_0} \frac{y'}{r'^3}  & \qquad
E_z = \gamma E'_z & = \gamma\frac{q}{4 \pi \epsilon_0} \frac{z'}{r'^3}  
\end{alignat*}

Expressing the particle primed coordinates
in terms of the coordinates in $S$, that is $x'=\gamma (x-vt)$, $y'$=$y$ and $z'$=$z$, the electric field components are 
\begin{align*}
E_x & = \frac{q}{4 \pi \epsilon_0} \frac{\gamma (x-vt)}{[\gamma^2 (x-vt)^2+y^2+z^2]^{3/2}} \\
E_y & = \frac{q}{4 \pi \epsilon_0} \frac{\gamma y}{[\gamma^2 (x-vt)^2+y^2+z^2]^{3/2}} \\
E_z & = \frac{q}{4 \pi \epsilon_0} \frac{\gamma z}{[\gamma^2 (x-vt)^2+y^2+z^2]^{3/2}}
\end{align*}

As the particle is moving, in $S$ there is also a magnetic field. Using the magnetic field transformations it is

$$
\vec B' =0 = \vec{B}_{\parallel}+\gamma_v\bigl(\vec{B}_\bot -\frac{\vec{V}}{c^2}
\times \vec{E}\bigr) 
$$
which means
\begin{align*}
\vec{B}_{\parallel} & = 0 \\
\vec{B_\bot } & = \frac{1}{c^2} \vec{v} \times \vec{E}
\end{align*}


We may evaluate the electric field at the time $t=0$ 
\footnote{At a different time $\bar t$ the fields take in $(x,y,z)$ the same values as in $(x-v\bar t,y,z)$ at $t$=0.}

$$
\vec{E} = \frac{q}{4 \pi \epsilon_0} \frac{\gamma \vec{r}}{[\gamma^2 x^2+y^2+z^2]^{3/2}}
$$
Denoting with $\theta$ the angle between the $x$-axis and
$\vec{r}$ and using the relationship

$$\gamma^2 x^2+y^2+z^2=\gamma^2r^2(1-\beta^2\sin^2\theta) $$
we get 
$$
\vec{E}=\frac{q}{4 \pi \epsilon_0}
\frac{1-\beta^2}{(1-\beta^2 \sin^2\theta)^{3/2}}\frac{\hat{r}}{r^2}
$$
The electric field is still radial and follows the $1/r^2$ law,
but has no more a spherical symmetry.
The magnetic field is perpendicular to the plane defined by
$\vec{r}$ and $\vec{v}$.

The exact knowledge of the particle fields in an accelerator is important for 
instance for designing diagnostics and computing wake-fields.
The situation is simplified when 
particles are ``ultra-relativistic''
that is their speed in the accelerator frame is almost $c$.
For $\beta \rightarrow$ 1 it is $\vec E \rightarrow$ 0, unless
$\theta$=90$^0$ or 270$^0$ where the field is enhanced by a factor $\gamma$.

\section{Forces between moving charges}
\begin{figure}[htb]
\centering
\rotatebox{0}{
\includegraphics*[width=42mm]{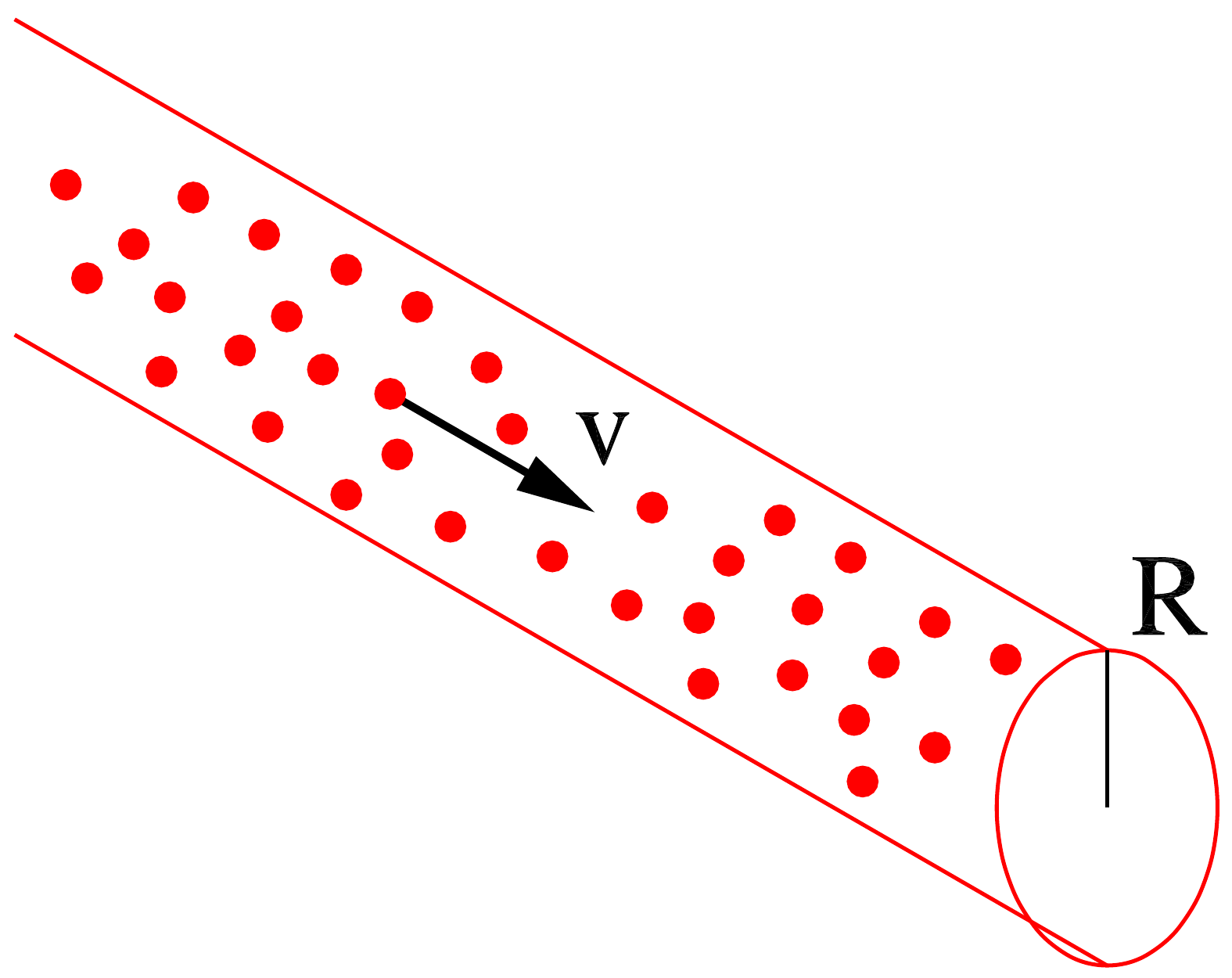}}
\caption{\label{eliana-rel-f27} Uniform cylindrical charge distribution.}
\end{figure}

Let us consider an uniform cylindrical beam of radius $R$
of equally charged particles
moving with velocity $\hat x v$  (see Fig.~\ref{eliana-rel-f27}).
Each of them experiences a repulsive
electric force and an attractive
magnetic force.

In the reference frame
$S'$ where the particles are at rest there is no magnetic
field. Inside the beam ($r'\le R$) it is

$$
F'_{r'} = q E'_{r'} = \frac{1}{2\pi \epsilon_0 R^2} {q^2 \lambda' r'}
$$
with $\lambda' = N/L'$ line density in $S'$.
The force acting on each charge is purely radial.

By using Eq.~(\ref{eliana-rel-eq16b})  for the Newton force transformation 
we get

\begin{align*}
F_\parallel = 0 \qquad
F_r = \frac{1}{\gamma} F'_{r'}
    = \frac{1}{2 \pi \epsilon_0 R^2} {q^2 \lambda r} \frac{1}{\gamma^2}
\end{align*}
where the line density in $S$, $\lambda$, is related to the line density in $S'$ by $\lambda  = \gamma \lambda'$.
In the reference frame $S$ the force is still radial and repulsive, but it is
reduced by a factor $1/\gamma^2$.

Beams in accelerators may be approximated by a uniform cylindrical charge distribution. 
We see that the repulsive force between the equally charged particles becomes smaller
at high energy.

\chapter{The CM energy}
The beams provided by accelerators to HEP collider or target experiments
allow  the creation of new particles and the study of the particles inner structure.
The energy
is not an invariant, however what matters is the energy in the center of mass which is 
therefore an
important parameter of a HEP facility.  

The center of momentum, usually referred as center of mass, for an isolated ensemble of particles  is defined
as the inertial frame where it holds
$$
\sum_i \vec p_i = \sum_i \frac{m_{0,i} \vec{v}_i}{\sqrt{1-V^2/c^2}}
                = 0
$$
where $V$ is the frame speed with respect to the laboratory.

We have seen that $(E/c)^2-|\vec p|^2$ is an invariant with value $m_0^2c^2$. 
For the total energy and momentum of the ensemble
$$E=  \sum_i E_i  \hspace*{6mm} \mbox{and} \hspace*{6mm} \vec P=\sum_i \vec p_i 
$$
the invariant evaluated in the CM frame is
\vspace*{-4mm}
$$
\biggl(\sum_iE_i/c\biggr)^2-\sum_i\vec{p}_i \cdot\sum_i\vec{p}_i
= \biggl(\sum_i E'_i/c\biggr)^2
$$
where $E'_i$ is the energy of the $i-th$ particle in the CM frame.
Let us consider two
simple cases:
\begin{itemize}
\item[a)] two ultra-relativistic particles colliding ``head-on'';
\vspace*{-2mm}
\item[b)] one ultra-relativistic particle hitting a particle at rest.
\end{itemize}

For the system of two particles it is
\begin{align*}
\frac{(E'_1+E'_2)^2}{c^2} & = \frac{(E_1+E_2)^2}{c^2}-(\vec{p}_1+\vec{p}_2)
                              \cdot (\vec{p}_1+\vec{p}_2) \\
                          & = \frac{(E_1+E_2)^2}{c^2}-p_1^2 -p_2^2 -2\vec{p}_1
                              \cdot \vec{p}_2
\end{align*}
Moreover for ultra-relativistic particles it is
$$
p = mv\simeq mc
  = \frac{E}{c}
$$

\begin{figure}[htb]
\centering
\rotatebox{0}{
\includegraphics*[width=42mm]{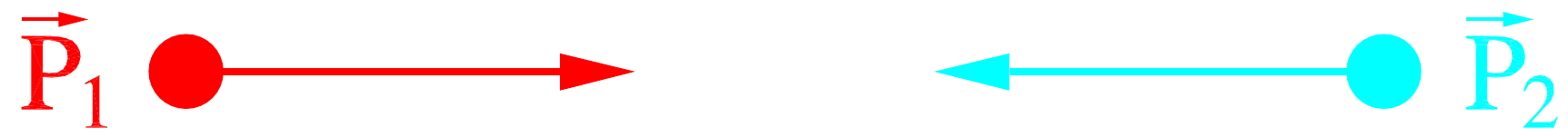}}
\caption{\label{eliana-rel-f28} Two particles colliding head-on.}
\end{figure}

Case a):
$\quad \vec{p}_1/p_1 = -\vec{p}_2/p_2$ (see Fig.~\ref{eliana-rel-f28}).

\vspace*{4mm}

\begin{gather*}
\frac{(E'_1+E'_2)^2}{c^2} = \frac{E_1^2}{c^2}+\frac{E_2^2}{c^2}
                           +2\frac{E_1 E_2}{c^2}-\frac{E_1^2}{c^2}
                           -\frac{E_2^2}{c^2}+2\frac{E_1 E_2}{c^2}
                          = 4\frac{E_1 E_2}{c^2} 
\end{gather*}  
              
and thus
$$ E'_1+E'_2 = 2\sqrt{E_1 E_2} $$

For instance, for the LHC $pp$ collider it is $E_1$=$E_2$=6.5 TeV and the energy in the center of mass is $E_1'+E_2'$=2$\times$6.5=13 TeV.
For the  $p/e^\pm$ HERA collider, which was in operation until 2007,
 with $E_1$=920 GeV and $E_2$=27.5 GeV it is $E'_1+E'_2$=318 GeV.
\vspace*{4mm}

\begin{figure}[htb]
\centering
\rotatebox{0}{
\includegraphics*[width=42mm]{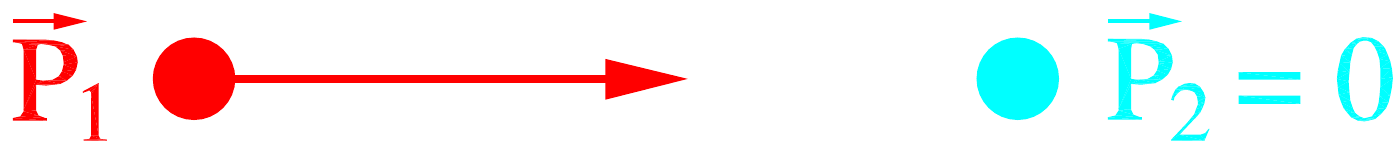}}
\caption{\label{eliana-rel-f29} Particle hitting a second particle at rest in the laboratory frame $S$.}
\end{figure}

Case b): $\quad \vec p_2 = 0$ and $E_2=m_{0,2}c^2$ (see Fig.~\ref{eliana-rel-f29}).

\vspace*{4mm}

\begin{align*}
\frac{(E'_1+E'_2)^2}{c^2} 
                          & = \frac{(E_1+E_2)^2}{c^2}-p_1^2 -p_2^2 -2\vec{p}_1
                             \cdot \vec{p}_2
\end{align*}                              
\begin{align*}
\frac{(E'_1+E'_2)^2}{c^2} & = \frac{E_1^2}{c^2}+\frac{E_2^2}{c^2}
                            +2\frac{E_1 E_2}{c^2}-\frac{E_1^2}{c^2}
                            = \frac{E_2^2}{c^2}+2\frac{E_1 E_2}{c^2} \\
(E'_1+E'_2) & = \sqrt{E_2(E_2+2E_1)}
              = \sqrt{E_2(m_{0,2}c^2+2E_1)}
              \simeq \sqrt{2 E_1 E_2}
\end{align*}
For example, with $E_2=0.938$~GeV (proton rest mass) to get in the
CM an energy of 318~GeV must be $E_1=$54~TeV.

From this example we see the advantage of collider experiments with respect to
fixed target ones in terms of available energy.

\chapter{The relativistic Hamiltonian of a particle in a EM field}

Let us consider 
a physical system in the presence of generalized forces
which can be derived from a function $U=U(q_i, \dot q_i)$ (\emph{generalized potential}).

It is possible to associate to such system a
lagrangian function $\mathcal{L}=T-U$, $T$ being the kinetic energy.
The dynamics of the system is described by the Lagrange equations
$$
\frac{d}{dt}\left(\frac{\partial \mathcal{L}}{\partial\dot{q}_j}\right)-
\frac{\partial \mathcal{L}}{\partial q_j} = 0
$$
The coordinates
$(q_{i},\dot{q}_{i})$
may be just the components of
$(\vec r_j,\dot{\vec r}_j)$, but it can be convenient
or even necessary to use other variables.

The generalized forces are related to
the function $U$ by~\cite{HG65}
\begin{align*}
F_\alpha & = \sum_i\frac{\partial q_i}{\partial r_\alpha}
             \left[-\frac{\partial U}{\partial q_i}+\frac{d}{dt}
             \frac{\partial U}{\partial \dot q_i}\right] 
\end{align*}             
which if the coordinates $(\vec r, \dot{\vec r})$ are used ($\partial q_i/\partial r_\alpha$=$\delta_{i\alpha}$) gives
\begin{align*}
F_\alpha & = -\left(\nabla U\right)_\alpha+\frac{d}{dt}
             \frac{\partial U}{\partial v_\alpha} \hspace*{6mm} \rightarrow \hspace*{4mm} \vec F = -\nabla U +
             \frac{d}{dt} \nabla_v U
\end{align*}

Very often physics problems are described
by using the Lagrange or the Hamilton formalism.
It is therefore useful
to derive the relativistic lagrangian function for a particle in an EM field.

The Hamilton principle says that between all patterns connecting
the point $(q_{i,1},\dot{q}_{i,1};t_1)$ to the point
$(q_{i,2},\dot{q}_{i,2};t_2)$
the system will actually follow that one for which the
integral (action)
$$
S = \int_{t_1}^{t_2} dt \mathcal{L}(q_i,\dot{q_i};t)
$$
has a minimum or a maximum.
This principle specifies the dynamics as well as
the Lagrange equations do.
First we find the lagrangian function for a \emph{free} particle,
$\mathcal{L}_{free}$, by asking
the action to be a Lorentz invariant.

We rewrite the action by using the proper time $d\tau=dt/\gamma$
$$
S = \int_{t_1}^{t_2} dt \mathcal{L}_{free}
  = \int_{\tau_1}^{\tau_2} d\tau\gamma \mathcal{L}_{free}
$$
In order to be $\gamma \mathcal{L}_{free}$ an invariant
$\mathcal{L}_{free}$ must be
proportional to $1/\gamma$ so that the dependence on $\gamma$ disappears
from the integral.

Let us write than $\mathcal{L}_{free}=\alpha/\gamma$.
For a free particle the Lagrange equation becomes
$$
\frac{d}{dt}\frac{\partial \mathcal{L}_{free}}{\partial v}  = 0 $$
Inserting our expression for $\mathcal{L}_{free}$ we have
$$
\frac{d}{dt}\frac{\partial}{\partial v}\frac{\alpha}{\gamma}
 = -\frac{1}{c^2}\frac{d}{dt}\alpha\gamma v
 = 0
 $$
which reduces to the Newton law $d(m_0\gamma v)/dt=0$ if we set
$\alpha=-m_0 c^2$.
Therefore the relativistic lagrangian function of the free particle is
$$
\mathcal{L}_{free} = -\frac{m_0c^2}{\gamma}
$$

\noindent
Now let us compute the lagrangian function related to the EM fields.
The Lorentz force is
$$
\vec{F} = q \vec{E} + q \vec{v} \times \vec{B} $$

The EM fields
in terms of scalar and vector potentials are (MKSA units)
\begin{gather*}
\vec{B} = \nabla \times \vec{A} \qquad\qquad
\vec{E} =-\nabla\Phi-\frac{\partial \vec{A}}{\partial t} 
\end{gather*}

Thus the Lorentz force can be written as
$$
\vec{F} = q\bigl(-\nabla\Phi-\frac{\partial \vec{A}}{\partial t}+
                \vec{v}\times\nabla\times\vec{A} \bigr) 
$$              
We use the identity
$$
\nabla(\vec{a} \cdot \vec{b})=\bigl(\vec{a} \cdot \nabla\bigr) \, \vec{b}+
                            \bigl(\vec{b} \cdot \nabla\bigr) \, \vec{a}+
                            \vec{a}\times \bigl(\nabla \times\vec{b}\bigr)+
                            \vec{b}\times \bigl(\nabla \times\vec{a}\bigr) 
$$
for transforming the term $\vec{v}\times\nabla\times\vec{A}$
$$
\vec{v}\times\nabla\times\vec{A}=\nabla \, \bigl(\vec{A} \cdot \vec v\bigr) -
                                 \bigl(\vec{v} \cdot \nabla\bigr) \, \vec{A}
$$

Thus the Lorentz force is
\begin{gather*}
\vec{F} = q\nabla\bigl(-\Phi + \vec{A} \cdot \vec{v} \bigr)
                       -q\frac{\partial \vec{A}}{\partial t}
                       -q\bigl(\vec{v} \cdot \nabla\bigr)\vec{A}
        = q\nabla\bigl(-\Phi + \vec{A} \cdot\vec{v}\bigr)
                       -q\frac{d\vec{A}}{dt} 
\end{gather*}                       
We recognize that the generalized potential is
$U=q\Phi -q\vec{A}\cdot \vec{v}$. Indeed
$$\frac{d}{dt} \nabla_v U = \frac{d}{dt} \nabla_v
                          \bigl(q\Phi -q\vec{A}\cdot \vec{v}\bigr)
                        = -q\frac{d\vec{A}}{dt}
$$
because the EM potentials
do not depend upon the particle velocity.

In conclusion, the Lorentz force for a particle in an EM field may be written in terms of a generalized potential $U$ as
$$
\vec{F} = -\nabla U + \frac{d}{dt} \nabla_v U $$
with 
$$U=q \, \bigl(\Phi - \vec{A} \cdot \vec{v}\bigr) $$
and the particle lagrangian related to the EM field is 
$$
\mathcal{L}_{int} = -U
                  = -q\Phi+q\vec{A}\cdot \vec{v}$$

The total lagrangian is obtained adding the
lagrangian of the free particle

\begin{equation}\label{eliana-rel-eq18}
\mathcal{L} = -\frac{m_0c^2}{\gamma}-q\Phi+q\vec{A}\cdot \vec{v}
\end{equation}

The hamiltonian function is related to the lagrangian function by

\begin{equation}\label{eliana-rel-eq19a}
\mathcal{H}(q_i,P_i) = \sum_iP_i\dot{q}_i-\mathcal{L}
\end{equation} 

with

$$P_i\equiv\frac{\partial \mathcal{L}}{\partial \dot{q}_i}=p_i+qA_i$$

The hamiltonian must be a function of $q_i$ and  $P_i$ and therefore 
we must express $\dot q_i$ (or $v_i$) in terms of $P_i$. From
$$ p_i= P_i-qA_i$$
and the relationship between momentum and energy
$$c^2p^2=E^2-E_0^2=m_0^2\gamma^2 c^4-m_0^2c^4$$
we get  
$$ m_0^2\gamma^2 c^2-m_0^2 c^2=p^2=(\vec P-q\vec A)\cdot (\vec P-q\vec A)$$
and therefore
$$
\vec{v}=\frac{\vec p}{m_0\gamma}
 = c\frac{\vec{P}-q\vec{A}}{\sqrt{m_0^2c^2+(\vec{P}-q\vec{A})^2}} 
 $$
Inserting this expression in Eqs.~(\ref{eliana-rel-eq18}) and (\ref{eliana-rel-eq19a}) we finally find the hamiltonian function
$$ 
\mathcal{H}(q_i,P_i) = \sum_iP_i\dot{q}_i-\mathcal{L}=c\sqrt{(\vec P-q\vec A)^2+m_0^2c^2} + q\Phi
$$

\chapter{Some relationships}

\begin{gather*}
\gamma \equiv \frac{1}{\sqrt{1-(v/c)^2}} \qquad\qquad
\beta \equiv \frac{v}{c}=\sqrt{1-\frac{1}{\gamma^2}} \\
m = \gamma m_0 \qquad
\vec{p} = \gamma m_0 \vec{v}
        = \frac{m_0 \vec{v}}{\sqrt{1-(v/c)^2}}
\qquad
\left(\frac{v}{c}\right)^2 = \frac{p^2}{(m_0c)^2+p^2 } \\
E = mc^2 \qquad\qquad
E_0 = m_0c^2 \qquad\qquad
\frac{E}{E_0} = \frac{m_0\gamma c^2}{m_0  c^2}
              = \gamma \\
T = E-E_0
  = m_0\gamma c^2-m_0c^2
  = m_0c^2(\gamma -1) \\
\begin{align*}
E^2 & = (T+E_0)^2
      = m^2c^4
      = m_0^2\gamma^2 c^4
      = \frac{m_0^2c^4}{1-(v/c)^2}
      = \frac{m_0^2c^4}{1-p^2/(m_0^2c^2+p^2)} \\
    & = \frac{m_0^2c^4}{m_0^2c^2}(m_0^2c^2+p^2)
      = m_0^2c^4+c^2p^2
\end{align*} \\
cp = c\gamma m_0 v
   = \frac{E}{E_0}c m_0 v
   = \frac{E}{m_0c^2}c m_0 v
   = \beta E
\qquad\qquad
cp\simeq E \quad \text{for} \quad \beta\rightarrow 1
\end{gather*}

A table of relationships between $\beta$, $\gamma$, momentum and relativistic energy, together 
with their relative variations, may be found in~\cite{BGGR70}.


\begin{thebibliography}{99} 

\bibitem{AE05} A.~Einstein, ``Zur Elektrodynamik bewegter K\"orper'', Ann. Physik, 17,  891 (1905).
English translation on the web at https://www.fourmilab.ch/etexts/einstein/specrel/www/.%
 \bibitem{RR68} R.~Resnick,
  ``Introduction to Special Relativity'',
  John Wiley \& Sons, 1968.
  \bibitem{HK72} J.~C.~Hafele, R.~E.~Keating, ``Around-the-World Atomic Clocks: Observed Relativistic Time Gains'',
   Science, Vol.~177, No.~4044 (Jul. 14, 1972), 166-168.
  \bibitem{DJ75} J. D. Jackson, ``Classical Electrodynamics'', John Wiley \& Sons, 1998.
\bibitem{LT09}  G.~N.~ Lewis and R.~ C.~Tolman,  ``Contributions from the Research Laboratory of Physical Chemistry of the Massachusetts Institute of Technology:  The Principle of Relativity, and Non-Newtonian Mechanics'', Proceedings of the American Academy of Arts and Sciences, 44, pp.709-726,1909.   See also on the web 
https://www.ias.ac.in/article/fulltext/reso/024/07/0729-0734.%
\bibitem{WB64} W.~Bertozzi,  American Journal of Physics, 32 (7): 551-555 (1964).
\bibitem{HH} H.~Henke, JUAS 2019 Lecture on Relativity.
\bibitem{HG65} {H.~Goldstein},
  ``Classical Mechanics'',
  Addison-Wesley, 1965.
  \bibitem{Landau} L.~D.~Landau and E.~M.~Lif\v sits, ``The classical Theory of Fields'', Pergamon, Oxford, 1962.
\bibitem{BGGR70}  C.~Bovet et al., ``A selection of formulae and data useful for the design of A.G. synchrotrons'', CERN-MPS-SI-Int-DL-70-4, on the web at http://cds.cern.ch/record/104153.
  
\end{thebibliography}
\end{document}